\documentclass[epj]{svjour}
\usepackage{latexsym}
\usepackage{graphics}
\usepackage[pdftex]{graphicx} 
\usepackage{amsmath}
\usepackage{empheq}
\usepackage{dcolumn}
\usepackage{bm}
\usepackage{color}
\usepackage{amsfonts}
\usepackage{cite}

\newcommand{\e}{ {\rm e}}
\newcommand{\ET}{ $\alpha$-(ET)$_2$I$_3$}

\newcommand{\bk}{ \bm{k}}
\newcommand{\bkD}{ \bm{k}_{\rm D}}
\newcommand{\eD}{ \epsilon_{\textsf{D}}}

\newcommand{\kx}{k_x}
\newcommand{\ky}{k_y}

\newcommand{\bX}{\bar{X}}
\newcommand{\bY}{\bar{Y}}

\newcommand{\g}{\gamma}

\newcommand{\tE}{\tilde{E}}

\newcommand{\eea}{E^{\rm SO}_{1}(\bk)}
\newcommand{\eeb}{E^{\rm SO}_{2}(\bk)}
\newcommand{\eec}{E^{\rm SO}_{3}(\bk)}
\newcommand{\eed}{E^{\rm SO}_{4}(\bk)}

\newcommand{\teeb}{\tilde{E}^{\rm SO}_{2}(\bk)}
\newcommand{\teec}{\tilde{E}^{\rm SO}_{3}(\bk)}

\newcommand{\eee}{E^{\rm SO}_{5}(\bk)}
\newcommand{\eef}{E^{\rm SO}_{6}(\bk)}
\newcommand{\eeg}{E^{\rm SO}_{7}(\bk)}
\newcommand{\eeh}{E^{\rm SO}_{8}(\bk)}

\begin{document}
\title{First-principles study of the effective Hamiltonian
 for Dirac fermions with spin-orbit coupling in two-dimensional molecular conductor $\alpha$-(BETS)$_2$I$_3$}
\author{Takao Tsumuraya\inst{1}\thanks{e-mail: tsumu@kumamoto-u.ac.jp}
 \and Yoshikazu Suzumura\inst{2}
}                     
%
%
\institute{Priority Organization for Innovation and Excellence, Kumamoto University, Kumamoto 860-8555, Japan 
\and Department of Physics, Nagoya University, Nagoya 464-8603, Japan}
\date{Received: 19 June 2020  / Accepted: 15 December 2020}
%
\abstract{
 We employed first-principles density-functional theory (DFT) calculations to characterize Dirac electrons 
 in quasi-two-dimensional molecular conductor $\alpha$-(BETS)$_2$I$_3$ [= $\alpha$-(BEDT-TSeF)$_2$I$_3$]
  at a low temperature of 30K. We provide a tight-binding model with intermolecular transfer energies evaluated from maximally localized Wannier functions, where the number of relevant transfer integrals is relatively large due to the delocalized character of Se $p$ orbitals. The spin-orbit coupling gives rise to an exotic insulating state with an indirect band gap of about 2 meV. We analyzed the energy spectrum with a Dirac cone close to the Fermi level to develop an effective Hamiltonian with site-potentials, which reproduces the spectrum obtained by the DFT band structure.
\PACS{
      {PACS-key}{describing text of that key}   \and
      {PACS-key}{describing text of that key}
     } 
} 
\maketitle
\section{\label{sec:level1}Introduction} 
~~~Graphene exhibits unique transport properties originating from an electronic state in which the valence and conduction bands touch at a discrete point on the Fermi level~($E_F$) and the band gap is zero.\cite{Novoselov2004, Neto2009}
This Dirac cone band structure exhibits linear dispersion in the low-energy region, which is described by a relativistic Dirac equation 
in two-dimensions. 
As a result, the electrons can move at high speed 
 as if they have no mass. 
Therefore, this emergent electronic state is called the massless Dirac fermion with a zero-gap state (ZGS).
However, cases in, which the discrete contact point (Dirac point) is located close to the $E_F$ are few. 

Using a tight-binding~(TB) model, Katayama {\it et al.} first identified a system in which 
the Dirac point is located on $E_F$ in the two-dimensional (2D) molecular conductor $\alpha$-(BEDT-TTF)$_2$I$_3$ under uniaxial pressure, where BEDT-TTF is bis(ethylenedithio)tetrathiafulvalene, hereafter referred to as ET.\cite{Kobayashi2004, Katayama2006_JPSJ75} 
This finding is based on 
model parameters calculated with the extended H\"{u}ckel method for an experimental structure.\cite{Kondo2005}
Furthermore, a first-principles calculation verified the Dirac cone.\cite{Kino2006}
Compared with graphene, the anisotropy of the molecular conductor gives a property associated with a tilted Dirac cone, which can be analyzed in terms of a 2 $\times$ 2 effective Hamiltonian.\cite{Kobayashi2007, Goerbig2008}

At ambient pressure, $\alpha$-(ET)$_2$I$_3$ exhibits 
metallic behavior above 135 K,\cite{Bender1984} while 
it becomes an insulator below 135~K, with charge ordering (CO) leading to a lack of inversion symmetry.\cite{Rothaemel1986, Kajita1992, Kino1995, SeoCO2000, Takano2001, Wojciechowskii03, Kakiuchi2007}
Interestingly, this insulating phase can be suppressed by applying both uniaxial and hydrostatic pressures and the ZGS emerges.
Under such pressure, nuclear magnetic resonance (NMR) measurements provide the clear evidence of the inversion symmetry.~\cite{Takahashi2010, Hirata2016, Katayama_EPJ}  
Importantly, the carrier mobility increases and the density decreases significantly when the sample is cooled from 300~K to 1.5 K.\cite{Kajita1992} 
This is explained by the ZGS, which shows 
 an almost temperature-independent  resistivity and the zero-mode Landau level.\cite{Tajima2000, Tajima_uniaxis2002, Landau2009}
 These findings have allowed the rapid progress of both experimental and theoretical studies for the molecular Dirac systems.\cite{Tajima2006_JPSJ75, Tajima2009_STAM10, Kobayashi2009_STAM10, Suzumura2012, Tajima2012, Kajita2014, Pddddt2Kato, Pddddt2Tsumu, Ptdmdt2Zhou} 

Recently, the selenium-substituted analog $\alpha$-(BETS)$_2$I$_3$ has attracted much attention as a candidate compound of ambient-pressure for bulk massless Dirac material [BETS = BEDT-TSeF = bis(ethylenedithio)tetraselenafulvalene]. 
At ambient pressure, $\alpha$-(BETS)$_2$I$_3$ also shows temperature-independent resistance above 50~K, but the temperature crossover 
 from the metal to insulator (M-I) of 50 K is lower than the CO transition temperature in $\alpha$-(ET)$_2$I$_3$.\cite{Inokuchi1995_BCSJ68}
However, the origin of the insulating state, specifically the presence or absence of CO transition at ambient pressure has yet to be clarified. 
The spin susceptibility at low temperature is quite similar between the $\alpha$-(BETS)$_2$I$_3$ and $\alpha$-(ET)$_2$I$_3$, 
and it remains necessary to identify the inversion symmetry, which would indicate the absence of CO in $\alpha$-(BETS)$_2$I$_3$.\cite{Takahashi2011_JPSJ80}

To this end, several experimental groups have recently examined the possibility of breaking the inversion symmetry at low temperatures using NMR\cite{Shimamoto2014} and synchrotron X-ray diffraction, recently.\cite{Kitou2020} 
In a previous study, one of the present authors took a theoretical approach, performing a first-principles density-functional theory (DFT) calculation for the experimental structures at ambient pressure.\cite{Kitou2020} 
A pair of anisotropic Dirac cones was found at a $general$ $\bk$-point.
The overall electronic structure is similar to that reported in a previous DFT study for the 0.7~GPa structure.~\cite{Alemany2012, Kondo2009}
Unlike TB calculations with H\"{u}ckel parameters, the Dirac cones we obtained are robust (i.e., not overtilted), and achieve the massless Dirac electron system.\cite{Kondo2009, Morinari2014}
As described above, unlike $ \alpha $-(ET)$_2$I$_3$, both experiments show that the inversion symmetry remains even below the M-I crossover temperature.
Accordingly, the bands may be in Kramers degeneracy, and the spin-orbit coupling (SOC) effect opens an indirect gap of $\sim$2~meV at the Dirac points [Fig.~\ref{fig1}]. 
The size of the band gap is generally consistent with the M--I crossover temperature of 50K, since (semi) local density approximation in DFT slightly underestimates the band gap. 
Furthermore, the $\mathbb{Z}_2$ topological invariants indicate a weak topological insulator, although that of the low-temperature CO phase in $\alpha$-(ET)$_2$I$_3$ is a trivial insulator.\cite{Kitou2020}

A reliable TB model is essential to comprehend the band of the Dirac electrons properly.\cite{Konschuh2010} 
However, an efficient method for extracting an effective TB model including SOC has not yet been fully established yet for molecular solids. 
When the SOC effect is weak, the transfer energies used in the diagonal element (i.e., the same spin ) have been calculated with the Wannier function in the absence of SOC.\cite{Sanvino2017} 
The off-diagonal matrix elements (opposite spin) were derived via the second variation with non-self-consistent full-relativistic band calculation\cite{Sanvino2017} or using complex transfer energies obtained from relativistic quantum chemistry calculation for two isolated monomers of BETS.\cite{Winter2017}
Although these approaches have the advantage of extracting the form of a Hamiltonian whether or not SOC is present,
 a comparison of the band structure with that 
 of the DFT shows an overestimation of the band gap.\cite{Winter2017} 
Therefore, in this work, we developed an effective Hamiltonian generated from Bloch functions obtained in self-consistent DFT calculations with full-relativistic pseudopotentials. 
We found that the diagonal elements also contain a significant component coming from the SOC effect on Se $p$ orbitals. 
As shown later, the off-diagonal elements cause an energy gap $\sim$2 meV.

Furthermore, we note that 
 compared with the electronic state of\ET, the eigenvalues close to the Dirac points are in a quite-narrow energy window. 
The Wannier fitting to DFT bands indicates that the number of relevant transfer integrals is large due to the delocalized character of Se $p$ orbitals in the BETS molecule. 
To overcome this problem, we introduce 
  site-potentials that reasonably estimate the spectrum of the  DFT eigenvalues at several time-reversal invariant momentum (TRIM), and propose a precise effective TB model for the insulating state in $\alpha$-(BETS)$_2$I$_3$.

This paper is organized as follows. 
In Sec.~\ref{sec:level2}, we discuss the electronic structure of $\alpha$-(BETS)$_2$I$_3$ at ambient pressure from first-principles calculations.   
The computational details and crystal structure are presented in Sec.~\ref{sec:level21}.
Section~\ref{sec:level22} describes an overall band structure with the Dirac cone formation. 
In Sec.~\ref{TB}, we present an effective TB model extracted from DFT bands using MLWFs.
Section~\ref{sec:level4} describes the insulating state with optimized
 site-potentials. 
Furthermore, DOSs and the local charge densities are shown 
 to compare the results of the TB model and those of DFT calculations. 
In Sec.~\ref{sec:CompNoSOC}, we compare our results with a non-SOC TB model
and discuss the present calculation.
Finally, we conclude with a summary in Sec.~\ref{sec:level5}. 
 
\begin{figure}[tb]
  \centering
\includegraphics[width=0.85 \linewidth]{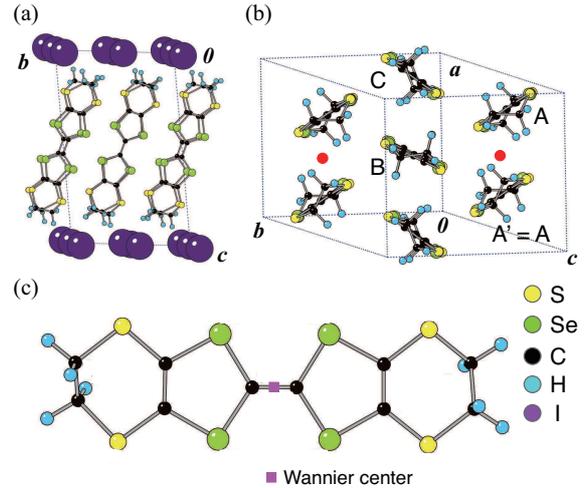} 
\caption{
Crystal structure of $\alpha$-(BETS)$_2$I$_3$ along (a) ${bc}$ and (b) ${ab}$ planes consisting of $A$, $A^{\prime}$, $B$, and $C$ molecules. 
Iodine atoms are not shown in this figure.
The molecular sites of $A$ and $A^{\prime}$ are crystallographically equivalent. 
The inversion center (red symbol) is located at the middle of A and A$^{\prime}$. 
(c) Molecular structure of BETS molecule. The Wannier center (solid square) is set at the center of C=C double bond of each molecule. 
}
\label{fig2}
\end{figure}

\begin{figure}[tb]
\begin{center}
  \includegraphics[width=1.0 \linewidth]{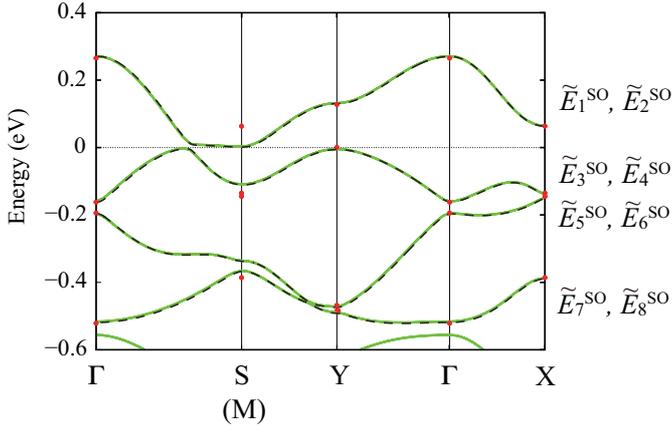}
\end{center}
\caption{(Color online)
Band structures including SOC effect along with the symmetric points in the first Brillouin zone, which denote time-reversal invariant momenta (TRIM) 
 given by 
 $\Gamma$ = (0,0,0),
 S~(M) = ($\frac{\pi}{\bf{b}}$, --$\frac{\pi}{\bf{a}}$, 0), Y = (0, --$\frac{\pi}{\bf{a}}$, 0), and X = ($\frac{\pi}{\bf{b}}$, 0, 0). 
The 2D vector is defined as $\bk$ = $k_1$$\bm{b^{\ast}}$ + $k_2$$\bm{a^{\ast}}$  = ($k_1$, $k_2$), 
where $\bm{b^{\ast}}$=($\frac{2\pi}{\bm{b}}$, 0, 0), $\bm{a^{\ast}}$=(0, $\frac{2\pi}{\bm{a}}$, 0).  
The solid and dashed curves are obtained by the first-principles DFT method and Wannier interpolation, respectively, which are explained precisely 
 in the last paragraph of Sec.~\ref{sec:CompNoSOC}. 
Solid circles on TRIM 
 are eigenvalues obtained from a tight-binding model 
based on the DFT and site-potentials shown in Table~\ref{Transfer_alpha}.  
The solid circles agree with those from the DFT calculations (solid curves) within an energy scale of 0.01 eV.  
The energy zero is set to be the top of the valence bands~[$\tilde{E}^{\rm SO}_{3}$($\bk$) and $\tilde{E}^{\rm SO}_{4}$($\bk$)].
}
\setlength\abovecaptionskip{0pt}
\label{fig3}
\end{figure}

\begin{figure}[tb]
\begin{center}
  \includegraphics[width=0.75 \linewidth]{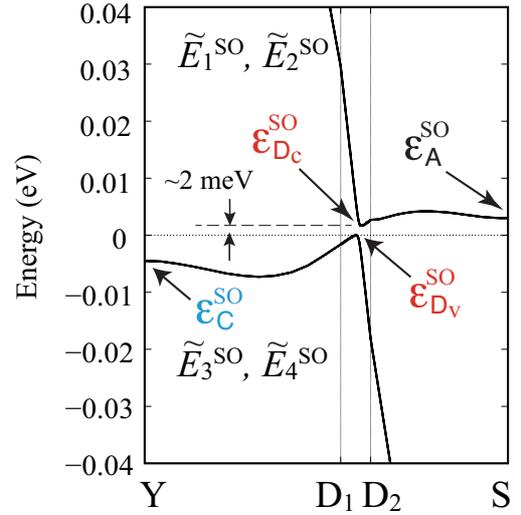}
\end{center}
\caption{(Color online)
First-principles band structure with SOC effect of $\alpha$-(BETS)$_2$I$_3$ at ambient pressure.\cite{Kitou2020} 
The energy dispersion of both conduction and valence bands ($E_c(\bk)$  and $E_v(\bk))$
  close to the Dirac cone is plotted along a line 
 including  D$_1$ ($k_1$, $k_2$) = (0.3095, --0.2995) 
    and  D$_2$ = (0.3595, --0.2995).
     Close to the Dirac point ($\bkD$),
 $\epsilon^{\rm{SO}}_{\rm A}$ 
 is 1.2 meV higher than that of $\textsf{D}_c$ 
 and  
 $\epsilon^{\rm{SO}}_{\rm C}$ 
 is 4.4 meV lower  than that of $\textsf{D}_v$ 
 where 
$\textsf{D}_v$ and $\textsf{D}_c$ denote the valence band  maximum and the conduction band minimum. 
\textsf{C} and \textsf{A} denote the momentum at (0, --0.5) and (0.5, --0.5), respectively. 
Note that 
 $\epsilon_{\textsf{D}_v}^{\rm{SO}} \simeq E_v (\bkD)$, and  
      $\epsilon_{\textsf{D}_c}^{\rm{SO}}\simeq E_c (\bkD)$. 
}
\setlength\abovecaptionskip{0pt}
\label{fig1}
\end{figure}

\section{\label{sec:level2}First-principles band structure} 
\subsection{\label{sec:level21} Calculation method and crystal structure}
~~~To derive low-energy effective Hamiltonians, we performed first-principles calculations based on DFT.\cite{HK1964, KS1965}
We used the generalized gradient approximation (GGA) proposed by Perdew, Burke, and Ernzerhof (PBE) as the exchange-correlation functional.\cite{GGAPBE} 
One-electron Kohn-Sham equations were solved self-consistently using a pseudopotential technique with plane wave basis sets adopting the projected augmented plane wave method,\cite{PAW1994} which was implemented in~\texttt{Quantum Espresso} (version 6.3).\cite{QE2009, QE2017} 
The cutoff energies for plane waves were set to be 55 (48) and 488 (488) Ry in the scalar (full) relativistic calculations, respectively. 
We used a 4 $\times$ 4 $\times$ 2 uniform $\bk$-point mesh with a Gaussian smearing method during self-consistent loops. 
For the calculations of the density of states (DOS), we used a uniform 18 $\times$ 18 $\times$ 2 $\bk$-point mesh. 
In both scalar and full relativistic pseudopotentials, the valence configurations were 1$s^1$, 2$s^2$2$p^2$, 3$s^2$3$p^4$, 4$s^2$4$p^4$3$d^{10}$, and 5$s^2$5$p^5$4$d^{10}$ for H, C, S, Se, and I atoms, respectively.
The pseudopotentials were generated using \texttt{atomic} code (version 6.3)\cite{atomicPP} with a pseudization algorithm proposed by Troullier and Martins.\cite{TM1991}
Using the Bloch wavefunctions obtained in the first-principles calculation described above, a Wannier basis set was constructed by using the \texttt{wannier90} code.\cite{Marzari1997, Isouza2001} 

The calculated crystal structures are based on an experimental structure measured at 30K,\cite{Kitou2020} for which structural optimization for the hydrogen positions was performed.
The crystal structure of $\alpha$-(BETS)$_2$I$_3$ is a triclinic structure with the space group of $P\bar{1}$,\cite{Kondo2009, Kitou2020} 
which is isostructural to the high-temperature phase of its sulfur analog, $\alpha$-(ET)$_2$I$_3$.\cite{Kondo2005, Bender1984, Kakiuchi2007} 
Figure~\ref{fig2}(a) shows the $bc$ planes where BETS molecules alternate with layers of iodine ions, I$_3$$^{-}$. 
In Fig.~\ref{fig2}(b), the BETS molecules form a herringbone pattern in the $ac$-plane.
The unit cell contains three crystallographically independent BETS molecules referred as to $A$~($A^{\prime}$), $B$, and $C$, where $A$ and $A^{\prime}$ molecules are connected by the inversion symmetry. 
The herringbone pattern is formed by two chains consisting of a layer in which $A$ (and $A^{\prime}$) molecules are stacked along the $a$-axis and another layer in which $B$ and $C$ molecules are stacked. 
Figure~\ref{fig2}(c) shows the structure of the BETS molecule, where Se atoms replace the central four S atoms connected with the central two C atoms in the ET molecule.

\subsection{\label{sec:level22} 
 Band structure of insulating state induced by spin-orbit coupling} 
~~~Figure~\ref{fig3} shows the calculated band structures close to the Dirac point including the SOC effect.
These band structures are plotted along a high-symmetric line of the first Brillouin zone. 
The overall band structure has many common features with the sulfur analog of $\alpha$-(ET)$_2$I$_3$.\cite{Kino2006}  
These bands are made up of a linear combination of the highest occupied molecular orbital~(HOMO), like the wavefunctions of the constituent BETS molecules.
This electronic structure's remarkable difference with the electronic state of $\alpha$-(ET)$_2$I$_3$ was discussed in a previous DFT study.\cite{Kitou2020} 
Four~(eight) bands near the $E_F$ occur in the absence (presence) of SOC in the energy range from --0.6 to +0.3 eV; these are attributed to the existence of four monomers in the unit cell.  
The band dispersions are referred to as $\tilde{E}^{\rm SO}_{1}$($\bk$), $\tilde{E}^{\rm SO}_{2}$($\bk$), $\cdots$ $\tilde{E}^{\rm SO}_{8}$($\bk$) in decreasing energy order.
However, as described in the introduction, two bands near the $E_F$ intersect along a line connecting the Y--D$_1$--D$_2$--S(M) points. 
This intersection creates a discrete contact point known as a Dirac point at $\bk_{\bf{D}}$ = ($k_1$, $k_2$) = ($\pm$0.333, $\mp$0.2995),
  which is located at $E_F$.  
When the SOC is considered, a small energy gap ($\simeq$ 2~meV) is opened close to the Dirac points as plotted in Fig.~\ref{fig1}.
 This is because the calculated structure is centrosymmetric, and every two bands [e.g. $\tilde{E}^{\rm SO}_{1}$($\bk$) and $\tilde{E}^{\rm SO}_{2}$($\bk$)] are Kramers degenerate. 
 The energy gap induced by SOC is an indirect band gap where  the wavenumber of the minimum of the conduction band is different from that of the maximum of the valence band.
In a previous study, TB Hamiltonian treated SOC as the second variation.\cite{Winter2017} Complex transfer energies were obtained by performing a quantum chemistry calculation overestimating the size of the band gap compared with that of the full-relativistic DFT calculations.  
The aim of the present study is to establish a scheme to derive effective Hamiltonian, including SOC, which can reproduce the DFT bands using MLWFs and site-potentials.

\section{\label{TB}$Ab$ $initio$ derivation of effective models}
\subsection{\label{sec:level3}Formulation for the tight-binding model}
~~~Based on first-principles calculations, the following two-dimensional model Hamiltonian was obtained  
\begin{subequations}
\begin{eqnarray}
H = \sum_{i,j = 1}^N \sum_{\alpha, \sigma} \sum_{\beta, \sigma'}
 t_{i,j; \alpha,\beta; \sigma, \sigma'} 
 a^{\dagger}_{i,\alpha, \sigma} a_{j, \beta, \sigma'} \; , 
\label{eq:Hij}
\end{eqnarray}
where $a^{\dagger}_{i, \alpha, \sigma}$ denotes a creation operator 
 of an electron 
 on each molecule $\alpha$~ 
 [= $A$, $A^{\prime}$, $B$, and $C$]  and  spin $\sigma$~$[= \uparrow,~\downarrow]$  
  in the unit cell 
  at  the $i$-th site of the square lattice. 
  The lattice constant is taken as unity.
The quantity  $t_{i,j; \alpha,\beta; \sigma, \sigma'}$ denotes 
 a  transfer energy defined by 
\begin{eqnarray}
t_{\alpha,\beta; \sigma, \sigma'}(\mathbf{R})=\langle\phi_{\alpha, \sigma,0}|H|\phi_{\beta, \sigma^{\prime},\mathbf{R}} \rangle ,
\label{equ1}
\end{eqnarray}
 where  $\phi_{\alpha, \sigma,\mathbf{R}}$ is the MLWF spread over the molecule $\alpha$  and centered at $\mathbf{R}$. 
To our knowledge, this is the first time that such transfer energy, including SOC effect, has been evaluated as is shown in the next subsection and listed in Table~\ref{Transfer_alpha}. 
We also examine the site-potential corresponding to the diagonal element 
 of Eq.~(\ref{equ1}), which is shown  in Appendix~A.
$H$ is the one-body part of the $ab$ $initio$ Hamiltonian. 
$\sigma$ and $\sigma^{\prime}$ are the index 
for spins $\uparrow$ and ~$\downarrow$.
  Equation (\ref{equ1}) shows that 
 $t_{i,j; \alpha,\beta; \sigma, \sigma'}$ 
   depends only on the difference between the $i$-th site and the $j$-th site.

After obtaining the Bloch functions using full relativistic DFT calculations, the Wannier functions were constructed using the \texttt{wannier90} code. 
To create the MLWFs, the eight bands near the $E_F$ shown in Fig. ~\ref{fig3} were selected as the low-energy degrees of freedom. 
Transfer energies are obtained from the overlaps between the eight (four) MLWFs in the presence (absence) of SOC. 
The center of each Wannier function is located at the middle of the central C = C bonds in each BETS molecule [solid square in~Fig.~\ref{fig2}(c)].
Using the Fourier transform
\begin{eqnarray}
a_{j,\alpha,\sigma} = 1/N^{1/2} \sum_{\bk}
      a_{\alpha,\sigma}(\bk) \exp[ i \bk \cdot \bm{r}_j], 
\label{eq:Fourier}
\end{eqnarray}
 Equation~(\ref{eq:Hij}) is rewritten as 
\begin{eqnarray}
 H =  \sum_{\bk} \sum_{\gamma, \gamma'} 
\hat{H}_{\gamma, \gamma'}(\bk) 
    a^{\dagger}_{\gamma}(\bk) a_{\gamma'}(\bk) \; , 
\label{eq:Hk}
\end{eqnarray}
where $\bk$ = $k_x$$\bm{b^{\ast}}$ + $k_y$$\bm{a^{\ast}}$  = ($k_x$, $k_y$) with 
$\bm{b^{\ast}}$=($\frac{2\pi}{\bm{b}}$, 0, 0), $\bm{a^{\ast}}$=(0, $\frac{2\pi}{\bm{a}}$, 0).   
We use ($k_x$, $k_y$) in stead of ($k_1$, $k_2$) in Fig. ~\ref{fig1}. 
In Eq.~(\ref{eq:Hk}), $\gamma= 1, 2, \cdots, 8$
 correspond to  
$A \uparrow, 
 A^{\prime} \uparrow, 
B \uparrow,   
C \uparrow,  
A \downarrow,   
 A^{\prime} \downarrow,   
 B \downarrow$, and 
 $C \downarrow$, respectively.
Using the intermolecular transfer energies shown in Figs. \ref{fig5} and \ref{fig6}, 
the  8 $\times$ 8 matrix including SOC 
 is obtained as   
\begin{eqnarray}
 \hat{H}(\bk) 
& = &
\begin{pmatrix}
  t_{11}   & t_{12}  & t_{13}  & t_{14} &   t_{15}   & t_{16}  & t_{17}  & t_{18}\\
  t_{21}   &  t_{22} & t_{23}  & t_{24} &   t_{25}   &  t_{26} & t_{27}  & t_{28} \\
  t_{31}   & t_{32}  & t_{33}  & t_{34} &   t_{35}   & t_{36}  & t_{37}  & t_{38} \\
  t_{41}   & t_{42}  & t_{43}  & t_{44} &   t_{45}   & t_{46}  & t_{47}  & t_{48}\\
  t_{51}   & t_{52}  & t_{53}  & t_{54} &   t_{55}   & t_{56}  & t_{57}  & t_{58}\\
  t_{61}   &  t_{62} & t_{63}  & t_{64} &   t_{65}   &  t_{66} & t_{67}  & t_{68} \\
  t_{71}   & t_{72}  & t_{73}  & t_{74} &   t_{75}   & t_{76}  & t_{77}  & t_{78} \\
  t_{81}   & t_{82}  & t_{83}  & t_{84} &   t_{85}   & t_{86}  & t_{87}  & t_{88}
\end{pmatrix} \; . 
\label{eq:H}
\end{eqnarray}
\end{subequations}
These matrix elements, $t_{ij} = (\hat{H})_{ij}$,
 are shown in Appendix B.
   From  Eq.~(\ref{eq:H}), energy bands $E_j^{\rm SO}(\bk)$ 
and wave function $\Psi_j(\bk)$ are calculated as 
\begin{subequations}
\begin{eqnarray}
\label{eq:eq5a}
 \hat{H} \Psi_j(\bk)
 &= & E_j^{\rm SO}(\bk) \Psi_j(\bk)  \; , \\
\Psi_j(\bk) &=&  \sum_{l} d_{j,l}(\bk)|l> \; .
\label{eq:eq5b} 
\end{eqnarray}
\end{subequations}
where $j = 1, \cdots, 8$, 
$l = A \uparrow, A'\uparrow, B \uparrow, \cdots, C \downarrow$, and 
$\Psi_j(\bk)$ denotes the corresponding wave function.
Note that $E_j^{\rm SO}(\bk)$ of the DFT calculation 
 should be distinguished.

In the following, we use  $\tilde{E}_j(\bk)=E_j(\bk)-\mu$,
 which is the decreasing  energy order.
In the absence of SOC,  the matrix elements becomes zero 
 for $t_{ij}$ with $i$=1, 2, 3, 4 and $j$= 5, 6, 7, 8.
 The corresponding energy band is given by 
 $E_1(\bk) = \eea = \eeb$, 
 $E_2(\bk) = \eec = \eed$, 
 $E_3(\bk) = \eee = \eef$, and 
 $E_4(\bk) = \eeg = \eeh$.
 
 We examined the Dirac electron between 
 the conduction band ($E_c$) and valence band ($E_v$), 
  which are given by 
$E_c = \eeb$ and $E_v = \eec$ for the presence of SOC 
and  
$E_c(\bk) = E_1(\bk)$ and $E_v(\bk) = E_2(\bk)$ for the absence of  SOC.
A Dirac point was obtained by $\bk = \bkD$   
  corresponding  to  a minimum of  $E_c(\bk)- E_v(\bk)$, 
 which becomes zero, i.e., $E_c=E_v=E_{\rm D}$ for  no SOC and  
finite  for SOC.
In the absence of SOC, the ZGS is obtained for $E_{\rm D} = \mu$,  ( $\mu$ being the chemical potential) and the semimetal is obtained for $E_{\rm D} \not= \mu$. 
In the presence of SOC, the insulating state is obtained for $\mu$ located in the gap between $\eeb$ and $\eec$. 

\begin{table}[b]
\centering
\caption{Effective transfer energies and energy differences in site-dependent potential energies ($V_B^{\rm{DFT}}$ and $V_C^{\rm{DFT}}$) in eV for $\alpha$-(BETS$_2$)I$_2$. 
$\Delta V_{B}$ and $\Delta V_{C}$ are the difference of site-potential energies of B and C molecular sites relative to $A$ (and $A^{\prime}$) site, respectively.  The definitions are shown in Appendix.~A.
}
\begin{minipage}{0.5\hsize}
\centering
\label{Transfer_alpha}
\begin{tabular}{crrrrrrr}
\hline\hline
$\sigma$ = $\sigma^{\prime}$ & SOC & Non-SOC \\ 
\hline
$a1$  & 0.0053 &  0.0058 \\
$a2$  & -0.0201 & -0.0197 \\ 
$a3$  & 0.0463 &  0.0471 \\ 
\hline        
$b1$  & 0.1389 & 0.1394 \\ 
$b2$  & 0.1583 & 0.1590  \\ 
$b3$  & 0.0649 & 0.0649  \\ 
$b4$  & 0.0190 & 0.0187  \\ 
\hline        
$a1'$ & 0.0135 & 0.0138  \\
$a3'$ & 0.0042 & 0.0043  \\ 
$a4'$ & 0.0217 & 0.0219  \\ 
\hline        
\hline        
$c1$  & -0.0024 & -0.0027 \\ 
$c2$  &  0.0063 & 0.0064  \\ 
$c3$  & -0.0036 & -0.0038 \\ 
$c4$  & 0.0013 & 0.0013  \\ 
\hline        
$d0$  & -0.0009 & -0.0009 \\ 
$d1$  & 0.0104 & 0.0104  \\ 
$d2$  & 0.0042 & 0.0042  \\ 
$d3$  & 0.0059 & 0.0059  \\ 
\hline        
$s1$  & -0.0016 & -0.0017 \\ 
$s3$  & -0.0014 & -0.0016 \\ 
$s4$  &  0.0023 & 0.0023  \\
\hline
\hline
$\Delta$$V_B^{\rm{DFT}}$ & -0.0047 & -0.0046 \\ 
$\Delta$$V_C^{\rm{DFT}}$  &  0.0208 &  0.0207 \\   
\hline\hline
\\
\end{tabular}
\end{minipage}
\begin{minipage}{0.45\hsize}
\centering
\begin{tabular}{cr}
\hline
\hline
$\sigma$ = --$\sigma^{\prime}$ & SOC\\ 
\hline
$b1_{so1}$  & -0.0020 \\ 
$b1_{so2}$  & 0.0020 \\
$b2_{so1}$  & -0.0019 \\
$b2_{so2}$  & 0.0019 \\ 
$b4_{so1}$  & -0.0008 \\ 
$b4_{so2}$  & 0.0008 \\ 
$c1_{so1}$  & 0.0007 \\ 
$c1_{so2}$  & -0.0007 \\ 
$c2_{so1}$  & 0.0003 \\ 
$c2_{so2}$  & -0.0003 \\ 
$c3_{so1}$  & 0.0006 \\
$c3_{so2}$  & -0.0006 \\
$c4_{so1}$  & 0.0001 \\ 
$c4_{so2}$  & -0.0001 \\  
\hline
\hline
\end{tabular}
\end{minipage}
\end{table}

\begin{table}[tb]
\caption{
Site-potentials and solutions for 
 $\alpha$-(BETS)$_2$I$_3$.
The unit for the potentials, $\textsf{A}$, $\textsf{D}$, and $\textsf{D}$ is
  in eV.
$n_A$, $n_B$, and $n_C$ are charge density localized on respective molecules. 
$\ast$ represents DFT or Optimized (Opt), which are DFT-derived site-potential defined in Appendix A and the optimized site-potential determined in Sec.\ref{sec:Opt_site_band}, respectively.
}
\label{table2}
\centering
\begin{tabular}{lrrrrrr}
\hline\hline
SOC & DFT & Opt \\ 
\hline 
$\Delta V_{B}^{\ast}$ & --0.0047 & --0.0047\\	
$\Delta V_{C}^{\ast}$ & 0.0208 & --0.0092	\\	
\hline
$\bkD$ & $\pm$(0.35, --0.29) & $\pm$(0.36, --0.29)\\
$\textsf{A}$ & 0.0038 & 0.0068	\\
$\textsf{D}_c$ & --0.0010 & 0.0006	\\
$\textsf{D}_v$ & --0.0028    & --0.0010	\\
$\textsf{C}$ & 0.0003 & --0.0024 \\	
$\mu$ & 0.1823 & 0.1684 \\
\hline
$n_A$(= $n_{A^{\prime}}$) & 1.48 & 1.46 \\
$n_B$ & 1.45 & 1.42 \\
$n_C$ & 1.59 & 1.65 \\
\hline
\hline
\end{tabular}
\end{table}

\subsection{\label{sec:level24}Transfer energies; the result of Wannier fitting}
\begin{figure}[tb]
\begin{center}
  \includegraphics[width=0.98 \linewidth]{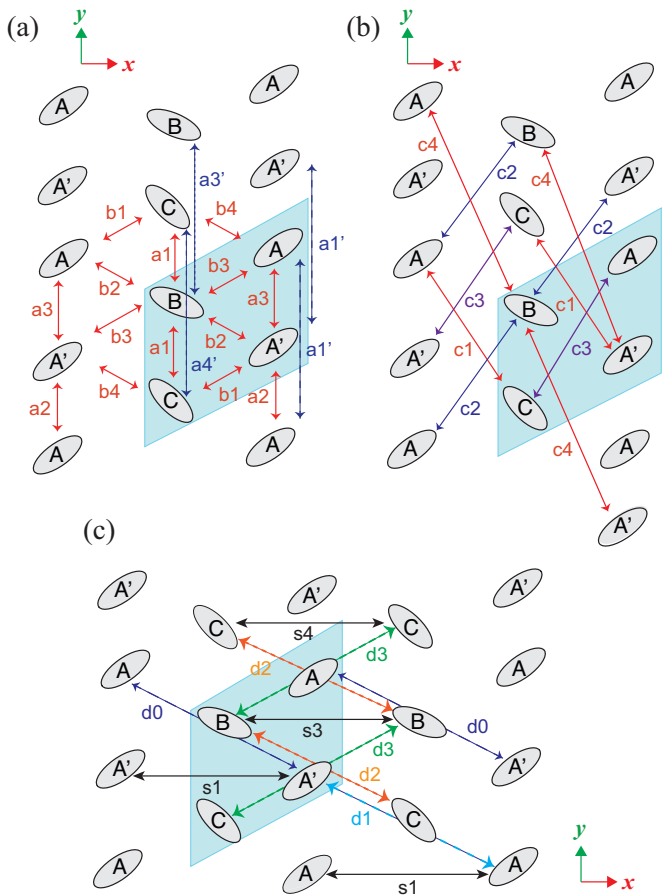}
\end{center}
\caption{(Color online)
Definition of intermolecular transfer energies $t_{i,j}$ of $\alpha$-(BETS)$_2$I$_3$. 
(a)~
$a_j$ and $b_j$ denote nearest neighbor transfer energies. 
$a_j'$ ($j$=1, 2, and 3) are transfer energies for the next nearest neighbors along the $a$-axis. 
These  notations refer to the previous works for $\alpha$-(ET)$_2$I$_3$.\cite{Kino2006}\cite{Mori_ET_1984} 
(b)~$c_j$ ($j$ = 1,... 4) are bonds along diagonal directions.  
 (c) $d_j$ ($j$= 0, 1, 2, and 3) are transfer energies hopping over one molecular-site along the same direction as $b_j$.
$s_j$ ($j$= 1, 3, and 4) are transfer energies between the same molecular sites in the next unit cell.
The parallelogram shown in the background is the home unit cell.
}
\setlength\abovecaptionskip{0pt}
\label{fig5}
\end{figure}

~~~Here, we detail how to evaluate the model parameters. 
The magnitude of transfer energies $t_{i,j; \alpha,\beta,\sigma = \sigma'}$ used in the diagonal element are larger than 0.001 eV and are listed 
 in Table~\ref{Transfer_alpha}.  
The threshold is determined by a requirement for reproducing DFT bands close to the $E_F$. 
Therefore, following intermolecular hopping, we added to the previously reported transfer integrals for $\alpha$-(ET)$_2$I$_3$ shown in Fig~\ref{fig5}(a).\cite{Kino2006, Mori_ET_1984}
The transfer integrals along diagonal directions shown in Fig~\ref{fig5}(b) are referred to as $c_j$ ($j$ = 1,... 4). 
In Fig. ~\ref{fig5}(c),
 the next nearest neighbor hopping along the $b_j$ direction is defined as $d_j$ ($j$= 0, 1, 2, and 3) and  
the  hopping between the same molecular sites in the neighboring unit cell 
 is defined as $s_j$ ($j$= 1, 3, and 4).
Note that the transfer energies between the inter-planes (along $c$-axis) are smaller than 0.001 eV; these are much smaller than those in the intra-plane ($ab$-plane). Thus, this system can be considered as a quasi-2D electron system. We note that the transfer energies used in the diagonal elements ($a_1$, $a_2$, $\cdots$ and $s_4$ on the non-SOC column listed in Table~\ref{Transfer_alpha})  are
similar to those of the 4 $\times$ 4 model in the absence of SOC. However, the values are not exactly the same as the non-SOC transfer energies. 
We find that a small difference of  $a_1$ -- $s_4$ between the presence and absence of SOC crucially changes the electronic state near $E_F$. We will discuss this point in Sec.~\ref{sec:band_TB}.

\begin{figure}[tb]
\begin{center}
\includegraphics[width=0.6 \linewidth]{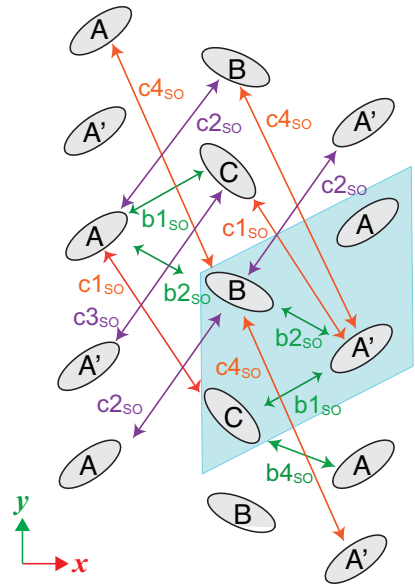} 
\end{center}
\caption{(Color online)
Definition of spin-orbit (SO) coupled transfer energies $b_{j}$$_{\rm{SO}}$ and $c_{j}$$_{\rm{SO}}$ of $\alpha$-(BETS)$_2$I$_3$.
Notations of the bonds of $b_j$$_{\rm{SO}}$ and $c_j$$_{\rm{SO}}$ follow those shown in Fig.~\ref{fig5}.
The parallelogram shown on the background is the home unit cell.
}
\setlength\abovecaptionskip{0pt}
\label{fig6}
\end{figure}

On the other hand, transfer energies used in the off-diagonal matrix element are obtained from overlaps between MLWFs with different spins $\sigma$.
The lattice structure in the molecular unit for $b1_{so1}, \cdots c4_{so2}$ for $\sigma$ = --$\sigma^{\prime}$ (i.e., the opposite spin) are shown in Fig. ~\ref{fig6}, and are referred to as spin-orbit~(SO) transfer energies. 
Here, the SO transfer energies are truncated at an absolute value of 0.0001 eV (Table~\ref{Transfer_alpha}). 
Interestingly, all the SO transfer integrals above the threshold are along diagonal directions whose bonds are $b_{j}$ and $c_{j}$, instead of $a_j$ and $a_j^{\prime}$.  

\subsection{\label{sec:level_A} DOS and local charge density at the molecular site}
~~We examined DOS and  local charge densities, which were obtained directly from the TB model 
 with   $\Delta V_{B}^{\rm{DFT}}$ = --0.0047 eV, 
   and $\Delta V_{C}^{\rm{DFT}}$ = --0.0012 eV (Table~\ref{fig2}). 
Using $E_j^{\rm SO}(\bk)$, the DOSs per site and per spin 
is obtained as
\begin{eqnarray}
\label{eq:DOS}
 D(\omega) = \frac{1}{2N} \sum_{j}\sum_{\bk}
     \delta (\omega - E_j^{\rm SO}(\bk)) \; . 
 \end{eqnarray}
Since the present system is a 3/4-filled band, the chemical potential  
$\mu$ is given by 
$ 3 = \int_{- \infty}^{\infty}  d \omega\; D(\omega) f(\omega -\mu) \; ,$
where $f(x) = 1/[\exp(x/T)+1]$ is the Fermi distribution function 
 with $T$ being the temperature absolute zero.
In terms of $d_{l}(\bk)$, 
 the local charge density, 
which denotes the electron number of each molecule per unit cell,
 is calculated as  
\begin{eqnarray}
\label{eq:n}
 n_{\alpha} 
 = \frac{1}{N}\sum_{j,\bk, \sigma}
  [d_{\alpha \sigma}^*(\bk) d_{\alpha \sigma}(\bk)]
     f( E_j^{\rm SO}(\bk) -\mu) \; . 
\end{eqnarray}
 with $\sigma = \uparrow, \downarrow$.

\section{\label{sec:level4}Dirac fermions: Insulating state with spin-orbit coupling}
\subsection{\label{sec:band_TB}Electronic state obtained from the effective models} 
\begin{figure}
  \centering
\includegraphics[width=4.2cm]{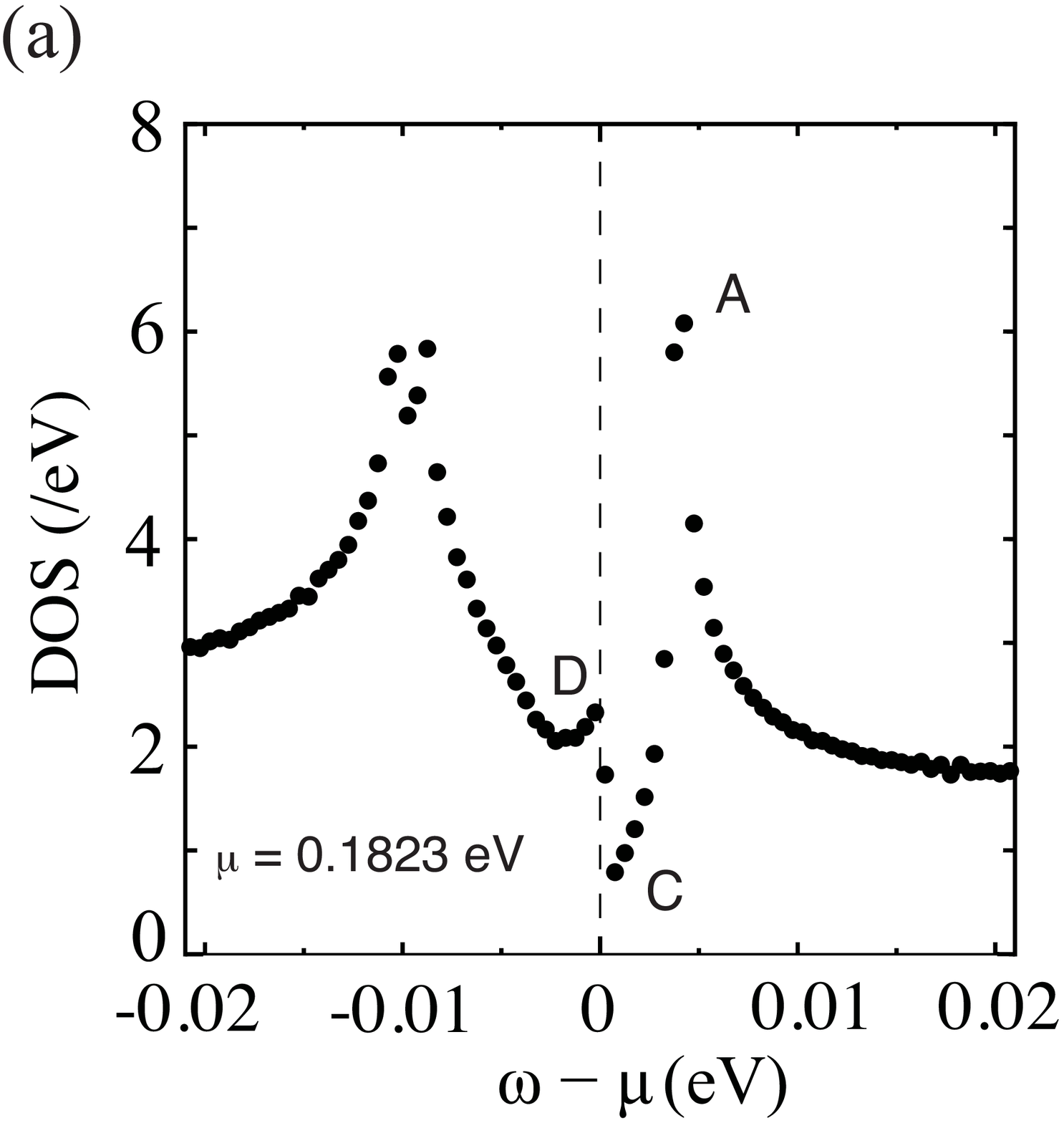}
\includegraphics[width=4.5cm]{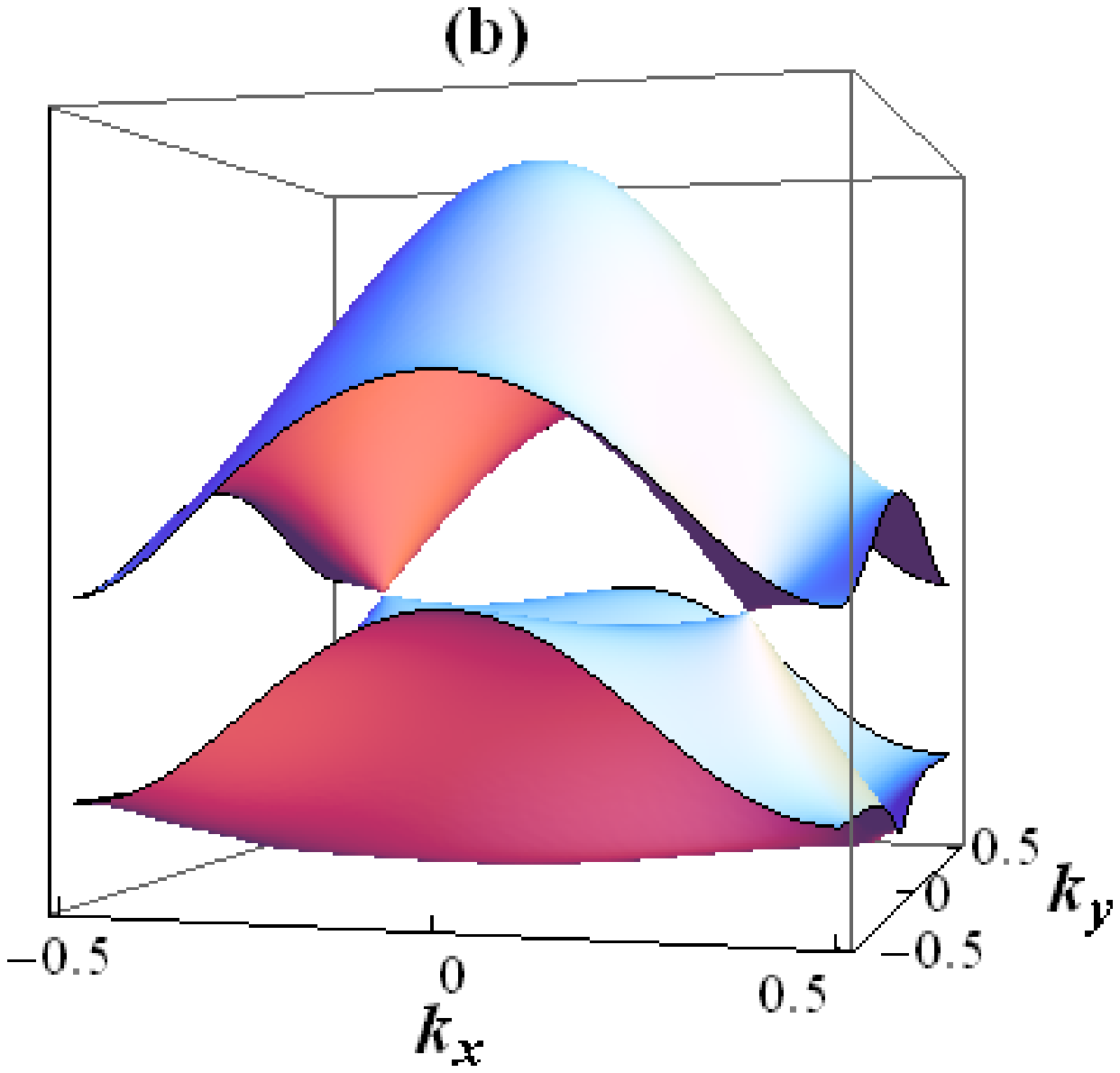} 
\caption{(Color online) 
(a) Density of states (DOS) with~SOC obtained from transfer energies and site-potentials shown in Table~\ref{Transfer_alpha}. 
The chemical potential $\mu$ locates at 0.1823 eV. 
The relative energies of the eigenvalues of $\textsf{A}$, $\textsf{C}$, and $\textsf{D}$ with respect to $\mu$ are 
$\textsf{A}$ = 0.0038 eV, $\textsf{C}$ = 0.0003 eV, 
 and $\textsf{D}$ = -0.0019 eV.
(b)~Two bands of $\eeb$ and $\eec$ 
  for the TB model with SOC. 
  These two bands are separated by a small gap of $\simeq$ 2 meV, 
 which is induced by SOC. 
The energy  of \textsf{D}, which is defined by a minimum of $\eeb - \eec$, is located at $\bk = \bkD = \pm (0.35, -0.29)$. 
}
\label{fig8}
\end{figure}

\begin{figure}
  \centering
 \includegraphics[width=4.5cm]{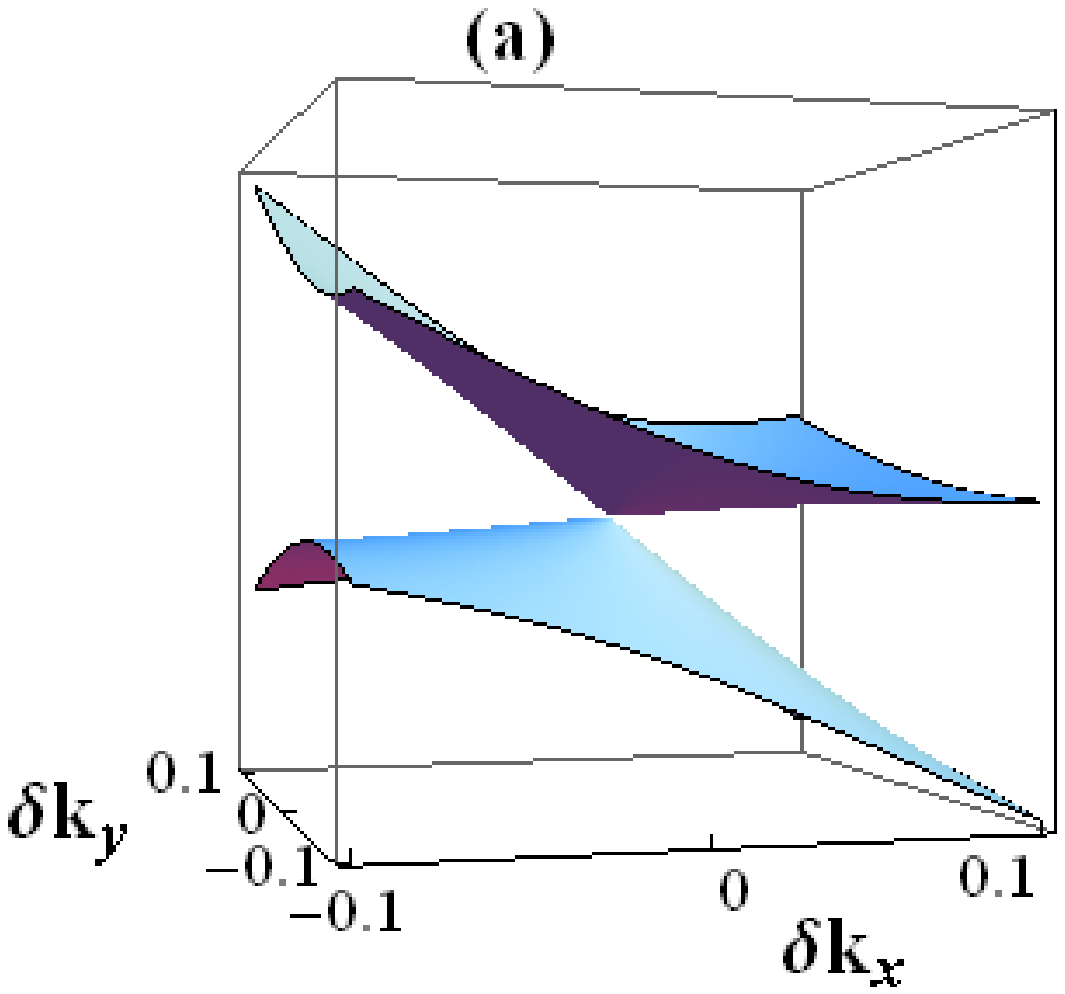}
 \includegraphics[width=3.5cm]{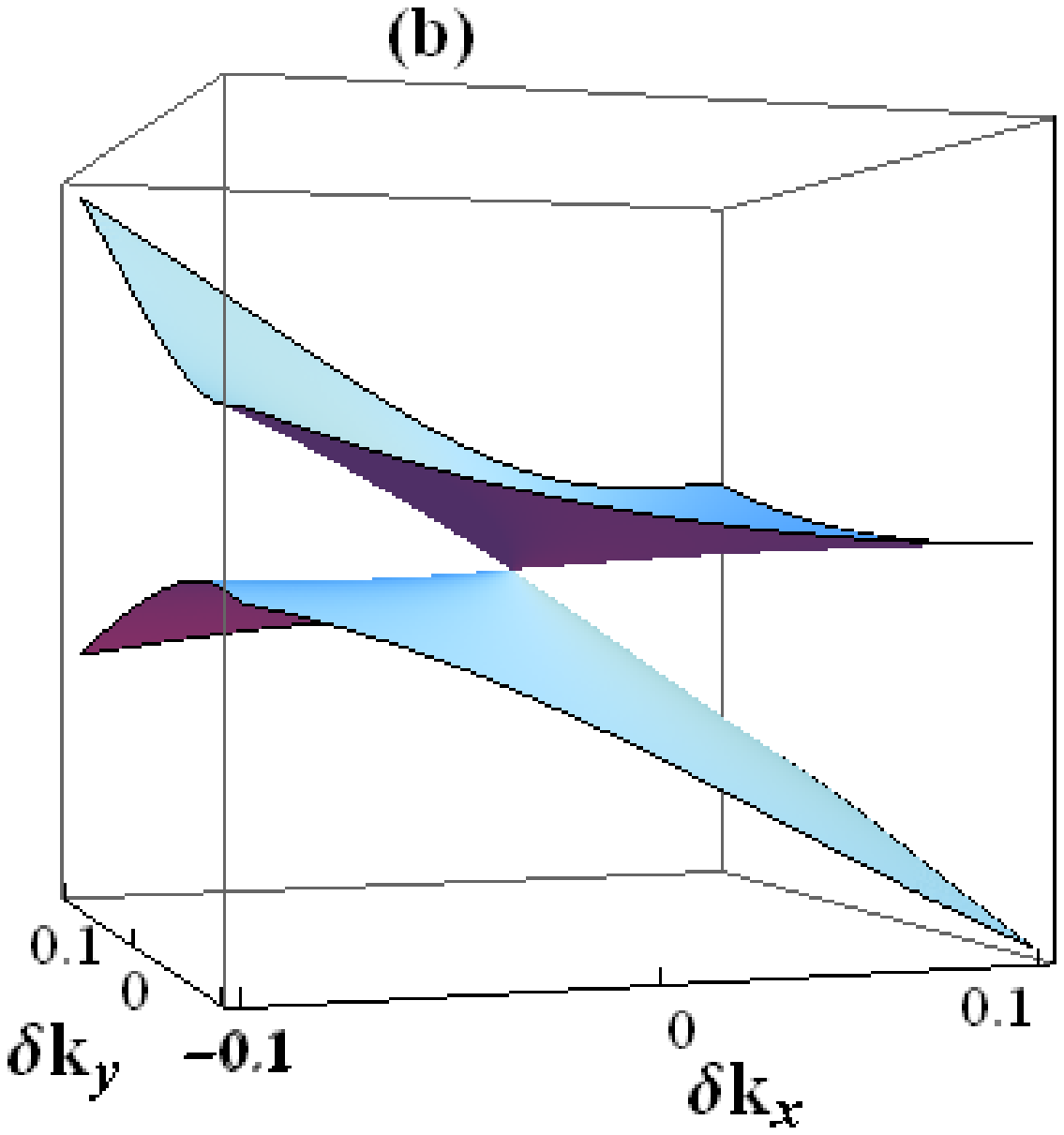}  
\includegraphics[width=4cm]{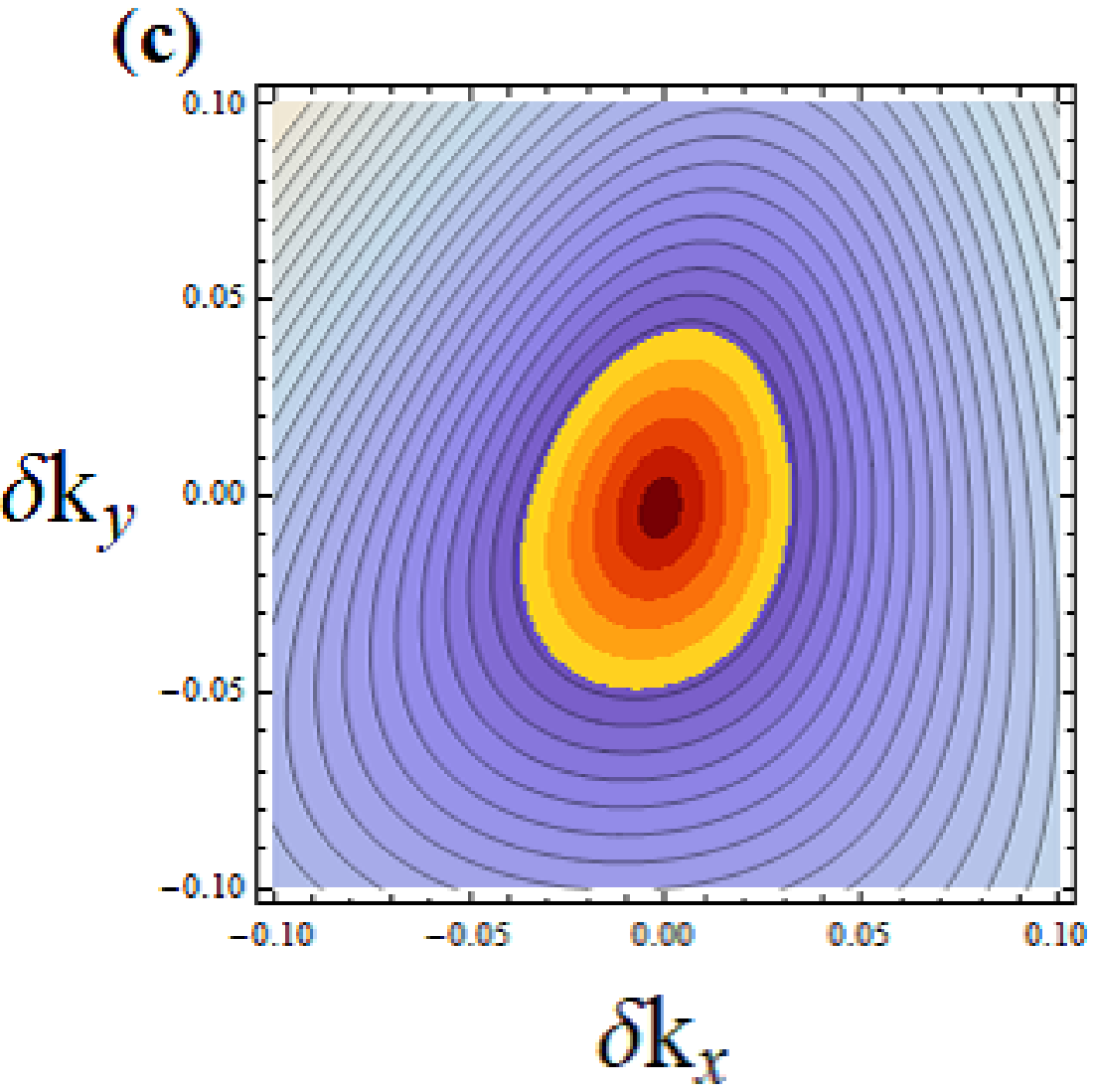} 
\includegraphics[width=4cm]{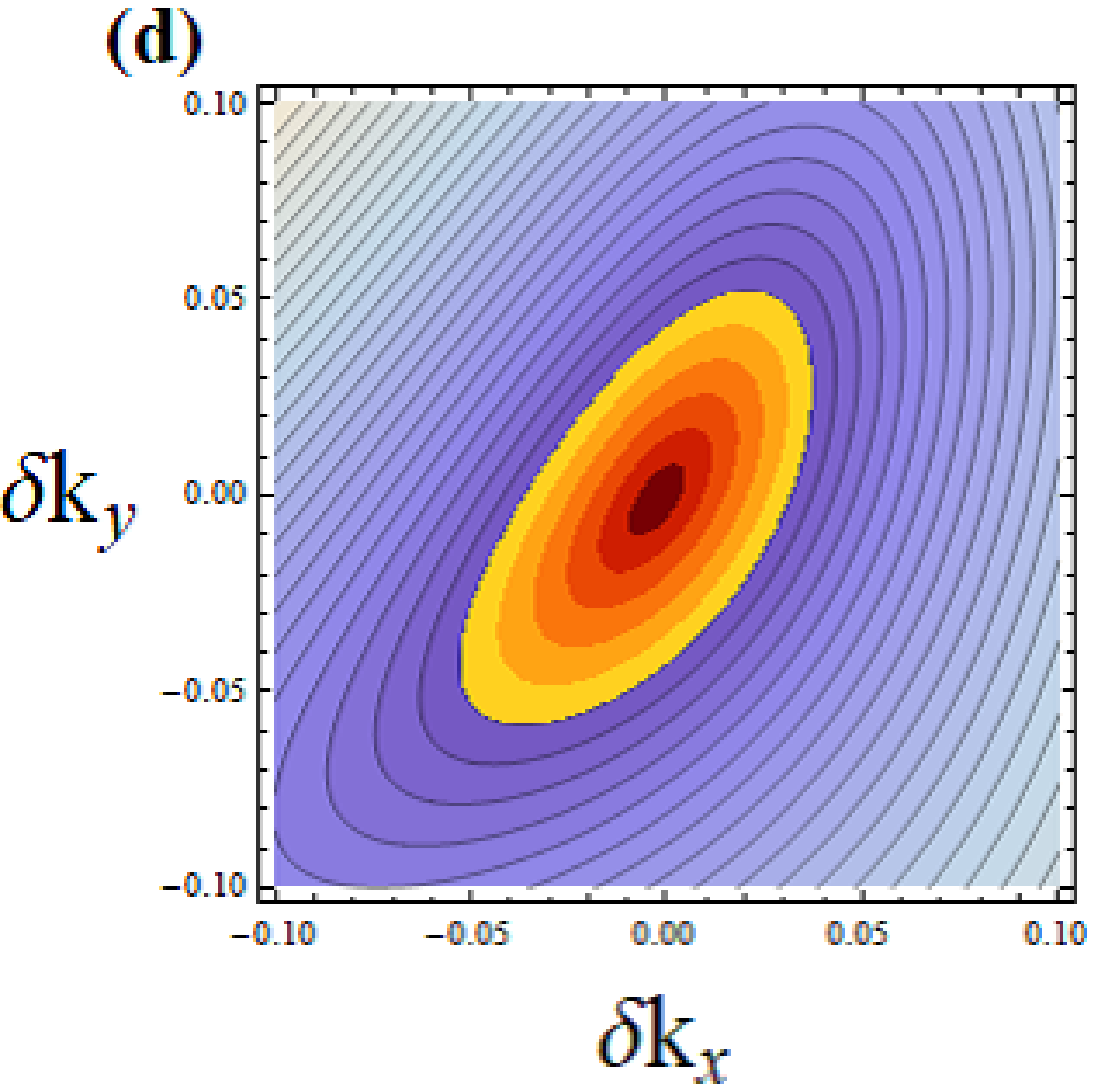} 
\caption{(Color online)
Dirac cone band structure (a) with SOC and (b) without SOC, 
on the plane of
  $\delta \bk (= \bk - \bkD)$ with  $\bkD = \pm(0.35,-0.29)$, which is
obtained from the effective model parameters shown in Table~\ref{Transfer_alpha}. 
The band gap  between  
two bands of $\eeb$ and $\eec$ is seen in (a), while 
two bands of $E_1(\bk)$ and $E_2(\bk)$ in (b) show 
 no gap.
(c) Contour plots of $\eeb - \eec $  as a function of $\delta \bk$,
 where the orange region corresponds to  $\eeb - \eec$ ($<$0.03 eV).
The quantity $\eeb - \eec$ takes a minimum 
  $\sim$ 0.0015 eV at $\bkD$.
(d) Contour plots of $E_1(\bk)$ $-$ $E_2(\bk)$ as a function of $\delta \bk$,
where the orange region corresponds to $E_1(\bk)$ $-$ $E_2(\bk)$ ($<$0.03 eV).
The quantity $E_1(\bk)$ $-$ $E_2(\bk)$ reaches a minimum 
  $\sim$ 0 eV at $\bkD$.
The anisotropy of the velocity of the cone of (d)    
 is large  compared with that of (c).
}
\label{fig81}
\end{figure}

~~~We first discuss the electronic structure obtained from the DFT-derived model parameters shown in Table~\ref{Transfer_alpha}.
As shown in Figs.~\ref{fig8}(a) and~\ref{fig8}(b), the present effective model with SOC does not reproduce the relative relation of the eigenvalues close to $E_F$ in the order of $\epsilon^{\rm{SO}}_{\rm C}$ $<$ $\epsilon^{\rm{SO}}_{\rm D}$ $<$ $\epsilon^{\rm{SO}}_{\rm A}$ in Fig. ~\ref{fig1}. 
Therefore, we made a small modification of the site-potentials to reproduce the DFT eigenvalues quantitatively (discussed in the next section). 
The details of the band structures close to the Dirac cone in the presence and the absence of SOC are shown in Fig.~\ref{fig81}(a) and~\ref{fig81}(b), respectively. 
As depicted in Fig.~\ref{fig81}(a), the model with 8 $\times$ 8 matrices opens the indirect band gap of 1.8 meV around the $k$-point where the Dirac point was located. 
The size of the energy gap correspond closely with those from the DFT calculations shown in Fig.~\ref{fig1}. 
The 4 $\times$ 4 model with the DFT-derived parameters also accurately reproduced the Dirac point~[Fig.~\ref{fig81}(b)]. 
Therefore, the SOC changes the electronic state from ZGS to a topological insulating state as discussed in a previous study.\cite{Kitou2020} 
However, the bulk system is nearly identical to the ZGS since the size of the gap 
due to the SOC is much smaller than the energy forming the Dirac cone.

As plotted in Fig.~\ref{fig3}, the eigenvalues shown in solid circles on the $\rm{\Gamma}$, Y, and X points, which is obtained from the parameters in Table~\ref{Transfer_alpha} agree with the DFT bands (solid curve) within an energy scale of 0.01 eV.
However, the eigenvalues at S point do not agree well with the DFT eigenvalues. 
When all the transfer energies are included in a TB model without the truncation of small transfer energies, 
the structure of Wannier interpolated bands (dashed curves) perfectly reproduces the DFT bands (solid curves). 
Small, distant transfer energies that we omitted from the present TB model are essential to account for such small energy differences between the DFT eigenvalues and those in the TB model. However, the eigenvalues close to the Dirac points occupy a very narrow energy window: $\epsilon^{\rm{SO}}_{\rm C}$ is 4.4 meV lower than that of the valence bands maximum close to the Dirac point $\textsf{D}_v$, and $\epsilon^{\rm{SO}}_{\rm A}$ is 1.2 meV higher than the conduction band minimum at the Dirac point $\textsf{D}_c$. 
To overcome this problem, we search values of site-potentials to reproduce the spectrum of Fig.~\ref{fig1} and the DOSs, providing a low-energy effective Hamiltonian with a moderate number of transfer energies. 
 
\subsection{\label{sec:Opt_site}Optimization of site-energy potentials} 
\begin{figure}
  \centering
\includegraphics[width=0.8 \linewidth]{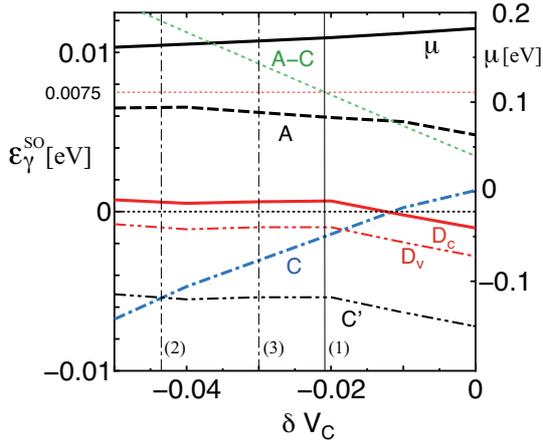}   
\caption{(Color online) 
$\delta V_{C}$ dependence of $\epsilon^{SO}_\gamma$ with  $\gamma$ = $\textsf{A}$, $\textsf{D}_c$, $\textsf{D}_v$, and $\textsf{C}$, 
  where $\textsf{D}_c$ and $\textsf{D}_v$ correspond to the conduction band minimum and valence band maximum at the Dirac point, respectively.
The dashed (black)  and dot-dashed (blue) lines denote the energy position of \textsf{A} and \textsf{C} obtained from the TB model, respectively, 
The  solid (red) and two-dots-dashed (red) lines denote 
  \textsf{D}$_c$ and \textsf{D}$_v$, respectively.
The crossing point given by the intersection of  $\textsf{A}-\textsf{C}$ fixed at 0.0075 eV 
 (from  DFT eigenvalues)  and  $\textsf{A}-\textsf{C}$ 
(the inclined  green dotted line) from the TB model gives   
  an optimized value  $\delta V_{C}$ = --0.022 eV  [line (1)].
Another  optimized value $\delta V_{C}$,  --0.043 eV [line (2)] 
 is also obtained from the intersection of 
$\textsf{C}$ and $\textsf{C}^\prime$, where 
$\textsf{C}^\prime$ = $\textsf{D}_v$--0.0044 eV
 is derived noting 
 the  difference between eigenvalues at $\textsf{C}$ and $\textsf{D}_v$ in Fig.~\ref{fig1}.
From lines (1) and (2), a compromise value  is taken as $\delta V_{C}$ = --0.03 eV [line (3)].
 }
\label{SOC_1}
\end{figure}

~~~To improve the spin-orbit Hamiltonian describing the insulating state, which is a novel state found at lower temperatures,
 we examined the site-potential,~\cite{Katayama_EPJ} which originates from a Hartree term of the Coulomb interaction treated within the mean-field consisting of the local density.
 We take  the chemical potential at the bottom of the conduction band near the Dirac cone, i.e.,  the minimum of $\eeb$.   
Using $\Delta$$V_B^{\rm{DFT}}$ and $\Delta$$V_C^{\rm{DFT}}$ as a reference, the site-potentials are rewritten as
\begin{eqnarray}
\Delta V_B = \Delta V_B^{\rm{DFT}} + \delta V_B, \\  \nonumber
\Delta V_C = \Delta V_C^{\rm{DFT}} + \delta V_C.  \nonumber
\end{eqnarray}
We examined to determine these site-potentials to reproduce the DFT eigenvalues of 
$\tE_2^{\rm{SO}}({\rm S})$ and $\tE_3^{\rm{SO}}({\rm Y})$
   in Fig.~\ref{fig1}. 
For this purpose, we developed the energy diagram shown in Fig.~\ref{SOC_1}. 
In this figure, we only show $\delta V_{C}$ dependence of $\textsf{A}$, $\textsf{D}$, and $\textsf{C}$ with the fixed $\Delta V_{B}$ = --0.0047 eV.
Note that during the exploration of site-potentials, we fixed the transfer energies as the DFT-derived parameters listed in Table~\ref{Transfer_alpha}. 
With decreasing $\delta V_{C}$,
$\textsf{C}$ decreases  while  $\textsf{A}$,
$\textsf{D}_v$ and $\textsf{D}_c$ increase.  
  For $\delta V_C < -0.02$ eV,  $\textsf{D}_v$ and $\textsf{D}_c$ becomes almost constant. 
(1) We first examine the variation of eigenvalue $\epsilon^{\rm{SO}}_\gamma$ ($\gamma$ = $\textsf{A}$, $\textsf{C}$, and $\textsf{D}$) by changing the site-potentials to correctly reproduce $\textsf{A} -\textsf{C}$ while maintaining the relation  $\textsf{C} <  \textsf{D}_v < \textsf{D}_c < \textsf{A}$ under a moderate choice of $\Delta V_{B}$. 
 Since $\textsf{A}-\textsf{C}$ calculated from first-principles is 0.0075 eV, 
the crossing point with $\textsf{A}-\textsf{C}$  is obtained  at  
 $\delta V_{C}$ = --0.022 eV [the line (1) in Fig.~\ref{SOC_1}].  
However, for this site-potential, the energy difference between $\textsf{C}$ and ${\textsf{D}_v}$ in the valence bands is much smaller than that 
 of the DFT  because the eigenvalue of $\textsf{A}$ is always higher than that in the DFT band. 
(2) In contrast to this method, we can  also determine the  site-potential using the relative energy difference between eigenvalues at $\textsf{C}$ and $\textsf{D}_v$ in Fig.~\ref{fig1}. Then, we newly define $\textsf{C}^\prime$ = $\textsf{D}_v$--0.0044 eV.
The crossing point between $\textsf{C}$ and $\textsf{C}^\prime$ is the solution. 
The optimized value of $\delta V_{C}$ is --0.043 eV [the line (2)]. 
On the contrary to (1), the valence bands close to the Dirac cone is well reproduced as shown in the inset of Fig.~\ref{SOC_4}(a). 
However, the depth of valley seen in the DOS near the $E_F$ is larger than that in the DFT calculation, since the energy position of $\textsf{A}$ is in higher energy. 
(3) We stress the situation, and chose a compromise value between these two solutions where $\delta V_{C}$ = --0.03 eV [the line (3)]. 
Based on Table \ref{Transfer_alpha}, site-potential $\Delta V_{C}$ is taken as a variational parameter given by $\Delta V_{C}$ = $\delta V_{C}^{\rm{DFT}}$ + $\delta V_{C}$ where $\Delta V_{C}^{\rm{DFT}}$ = 0.0208 eV. 
Hereafter, we use the following values: $\Delta V_{B}$ = --0.0047 eV, and $\Delta V_{C}$ = --0.0092 eV. 

\subsection{\label{sec:Opt_site_band}Electronic structure with the improved site-potentials} 
~~~The band structure with the optimized site-potentials of $\Delta V_{B}^{\rm{Opt}}$ = --0.0047 eV and $\Delta V_C^{\rm{Opt}}$ = --0.0092 eV is plotted in Fig.~\ref{fig7_1}(a).
This figure represents two bands of $\eeb$ and $\eec$, where the chemical potential is given by $\mu$ = 0.1684 eV.  
 $\eeb - \eec$ reaches a minimum at 
 $\bk = \bkD = \pm (0.36, -0.29)$, which is defined as a Dirac point in the case of SOC.
The energies of TRIM close to $\eD$  are  
 $\epsilon^{\rm{SO}}_{\textsf{A}} = \teeb$ at the S (=M) point and
 $\epsilon^{\rm{SO}}_{\textsf{C}} =\teec$ at the Y point.
Thus we obtained the following relationship:
\begin{eqnarray}
\epsilon^{\rm{SO}}_{\textsf{C}} < \epsilon^{\rm{SO}}_{\textsf{D}_v} < \epsilon^{\rm{SO}}_{\textsf{D}_c} < \epsilon^{\rm{SO}}_{\textsf{A}}. \nonumber
\end{eqnarray}

We depicted a contour plot of $\eeb - \eec$ in Fig.\ref{fig7_1}(b). 
A pair of Dirac points is found in the orange region, which is given by  
  0 $<$ $\eeb$ -- $\eec$ $<$ 0.03 eV. 
TRIMs in decreasing order of  $\eeb$ 
are given by  $\rm{\Gamma}$, Y, X, and S points.
Figure \ref{fig7_1}(c) shows 
 contour plots of $\teeb$ as a function of $\bk$, 
 where the cone is tilted toward the S. 
The orange region is given by 0 $<$ $\teeb$ $<$ 0.01 eV, which includes the S point 
 corresponding to the saddle point.
Figure \ref{fig7_1}(d) shows contour plots of $\teec$.
 The tilt of the Dirac cone is opposite to that of Fig.\ref{fig7_1}(c). 
 The orange region is described by --0.01 $<$ $\teec$ $<$ 0 eV, where 
there is a saddle point between  the Y and Dirac points.

\begin{figure}[tb]
   \centering
\includegraphics[width=0.98 \linewidth]{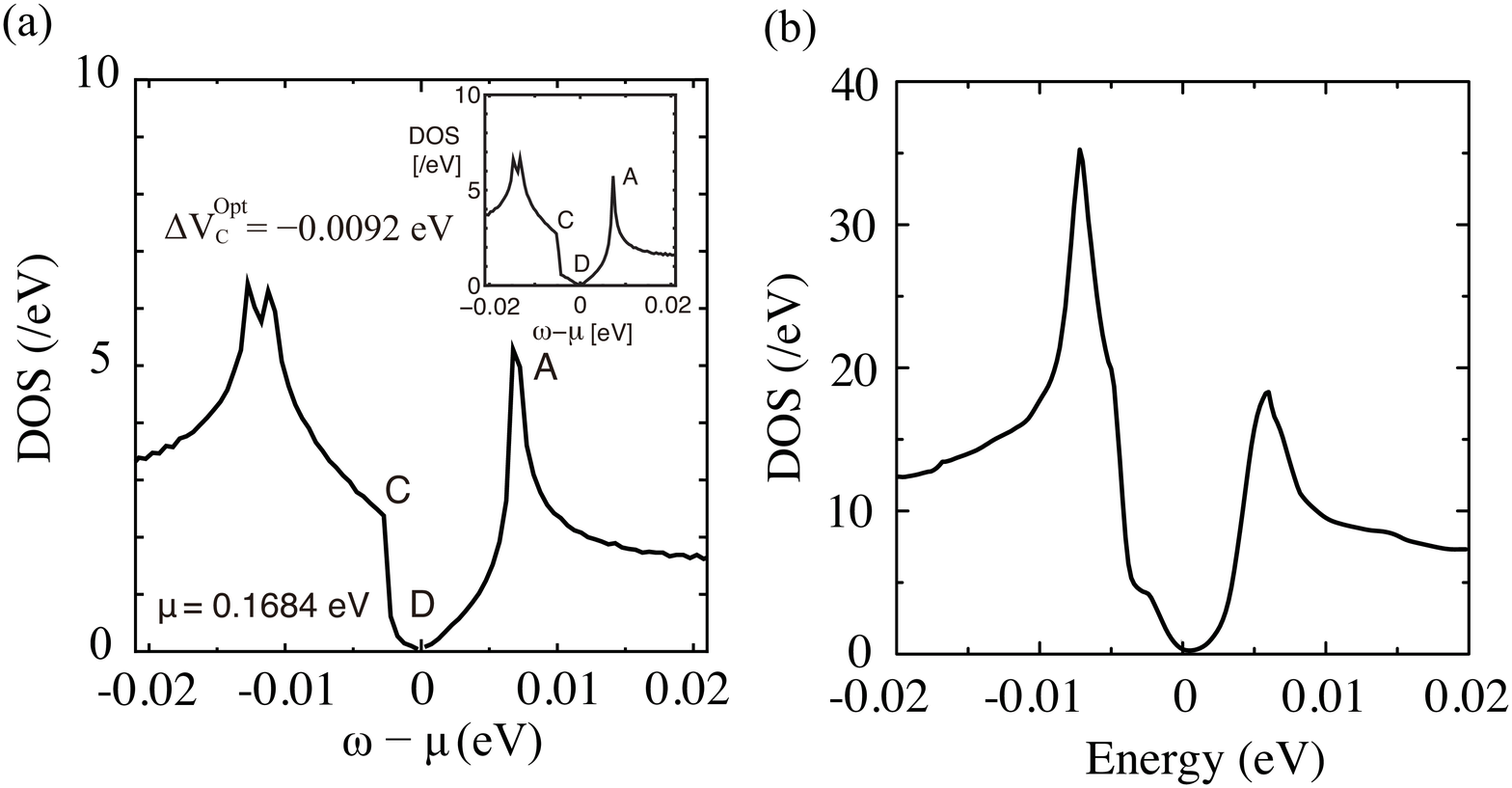}  
\caption{
(a)~Density of states (DOSs) per site and per spin  for a TB model with SOC as a function of $\omega - \mu$, where the site potential is taken 
 as  $\delta V_{C}$ = --0.03 eV (the line (3) of Fig.~\ref{SOC_1}) 
 with  $\mu$ = 0.1684 eV. 
The inset denotes DOS obtained for  $\delta V_{C}$ = --0.043 eV 
 (the line (2) of Fig.~\ref{SOC_1}) with $\mu$ = 0.1637 eV. 
 The symbols $\textsf{A}$,  $\textsf{C}$  and $\textsf{D}$ correspond to $\epsilon^{\rm{SO}}_{\textsf{A}}$, $\epsilon^{\rm{SO}}_{\textsf{C}}$, and $\epsilon^{\rm{SO}}_{\textsf{D}}$ (\textsf{D} is located between 
 $\textsf{D}_c$ and $\textsf{D}_v$), respectively.
(b) DOSs with SOC  directly calculated from the first-principles method. The zero of the energy is taken at the chemical potential (Fermi level).
The DOSs calculated from first-principles are also reported in Ref~\cite{Kitou2020}.
}
\label{SOC_4}
\end{figure}

\begin{figure}[tb]
  \centering
\includegraphics[width=4.2cm]{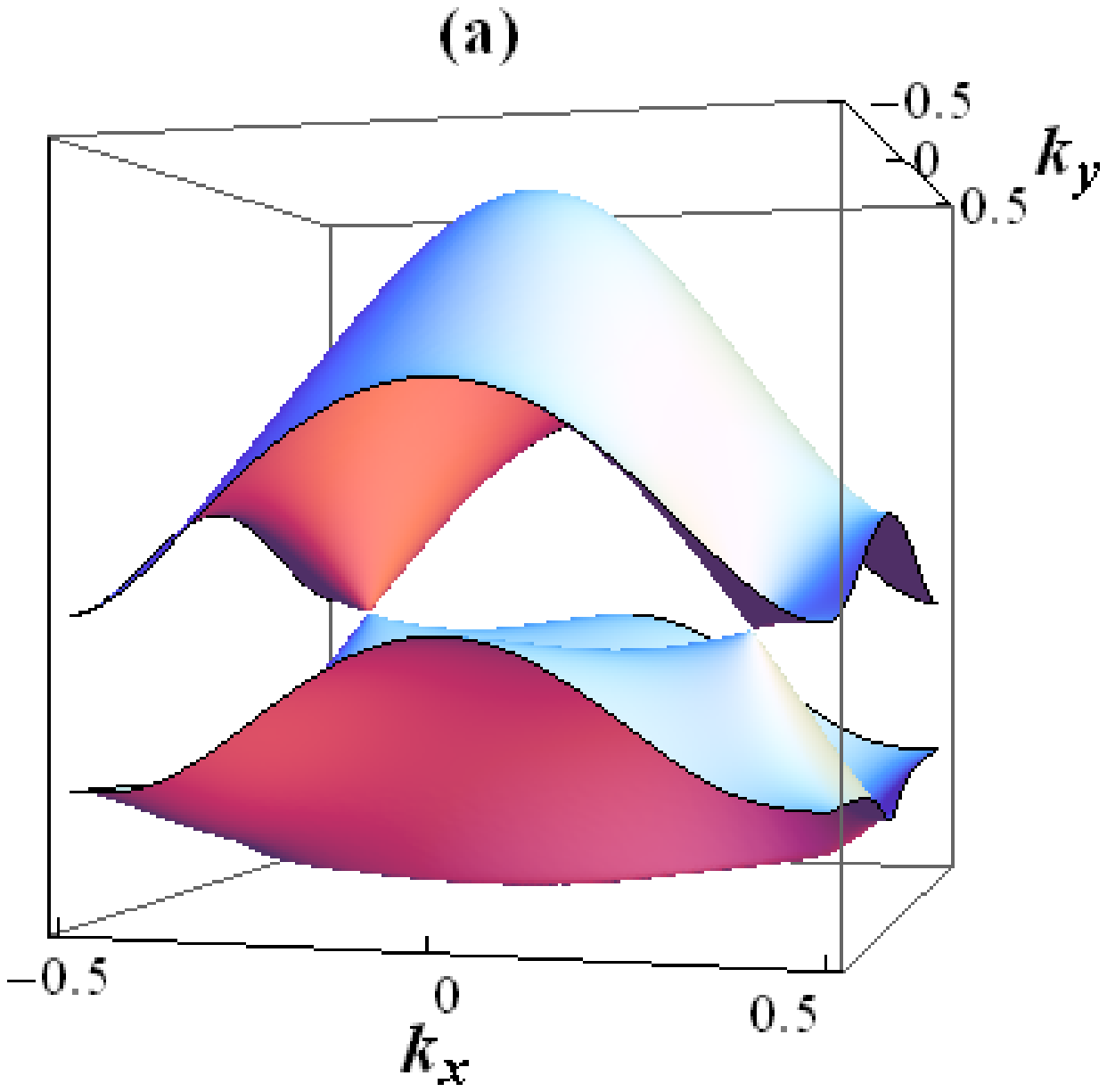} 
\includegraphics[width=4.2cm]{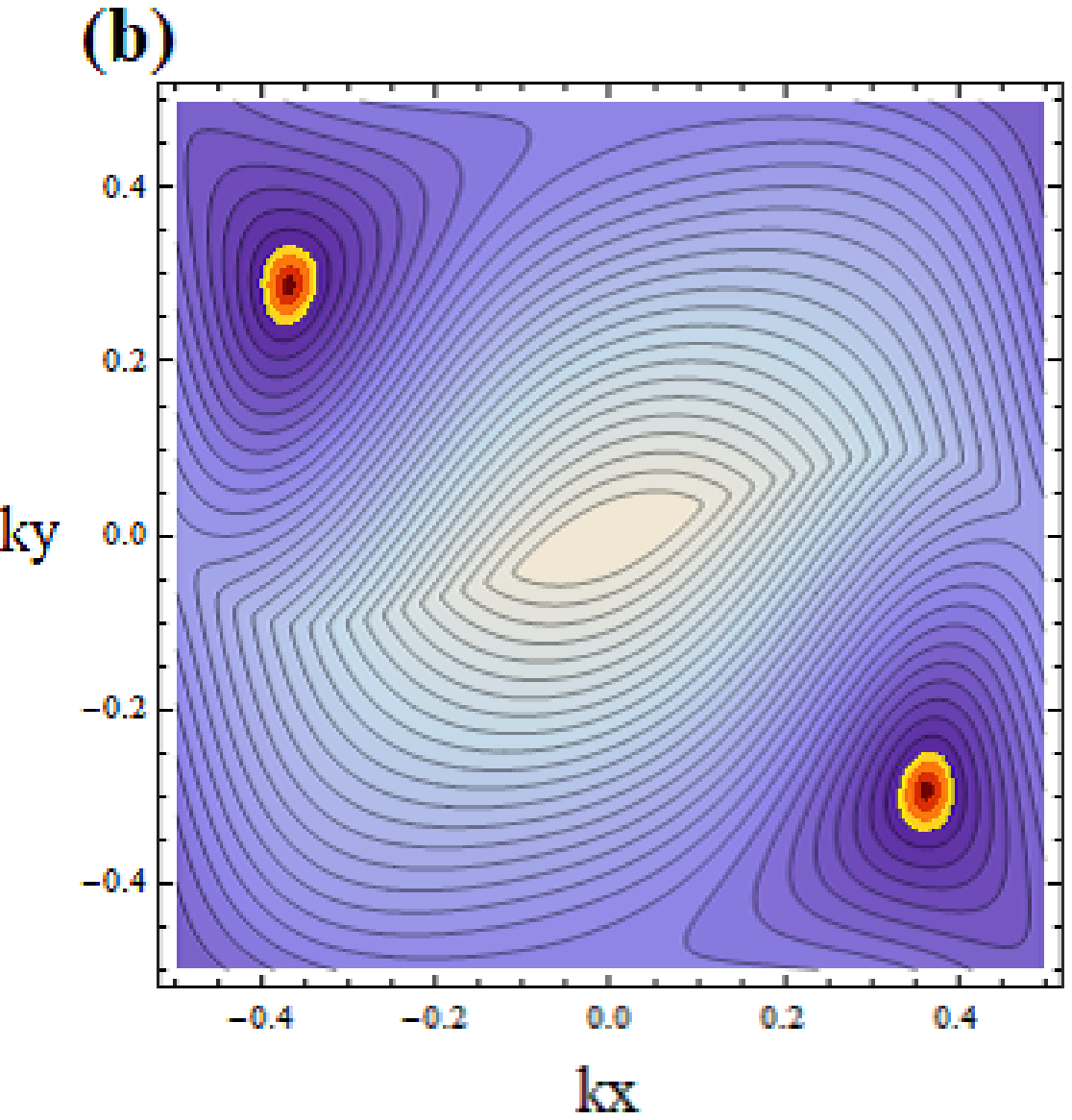} \\ 
\includegraphics[width=4.2cm]{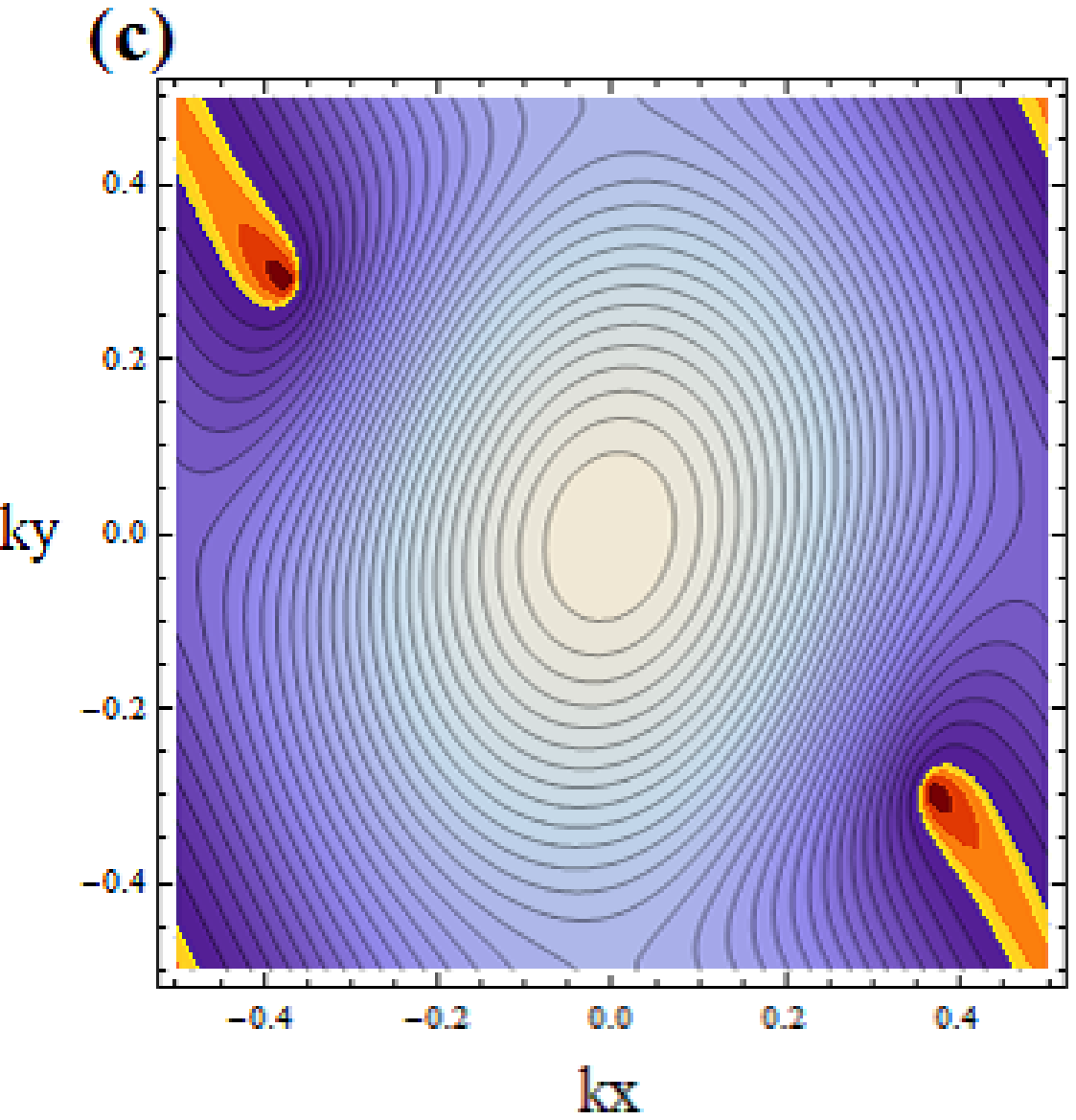} 
\includegraphics[width=4.2cm]{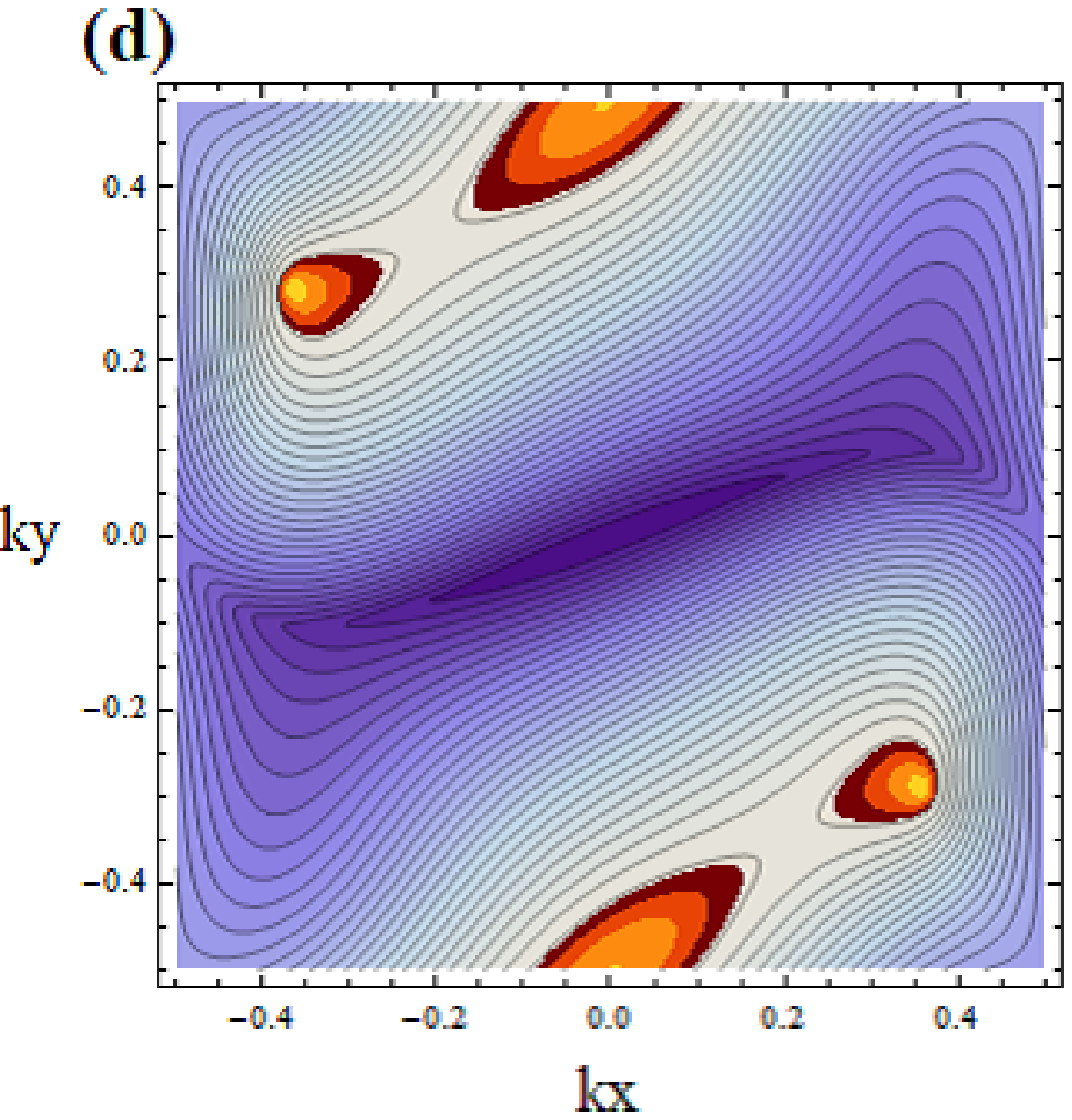} 
\caption{(Color online)
(a) Two bands  $\eeb$ and $\eec$ obtained from the TB model with SOC with the optimized site-potentials.
The chemical potential is given by $\mu$ = 0.1684 eV.  
The Dirac point, which is defined by 
 a  minimum of $\eeb - \eec$, 
   is  located at  $\bk$ = $\bkD$ = $\pm$(0.36, --0.29)
 with $\eD \simeq$ 0.
There is an insulating gap at $\sim$ 0.001 eV that is invisible in the figure.
(b) Contour plots of  $\eeb - \eec$. 
A pair of Dirac points is found in the orange region, which is obtained by  $0 < \eeb -\eec < 0.03$ eV. 
The Dirac point is located far from the TRIM. 
(c) Contour plots of $\teeb$ as the function of $\bk$, 
 where the cone is tilted. The orange region is given by 0 $<$ $\teeb$ $<$ 0.01 eV.
(d) Contour plots of $\teec$.
 The orange region is given by --0.01 eV $<$ $\teec$ $<$ 0 eV. 
 There is a  saddle point on a line  
 connecting Y and Dirac points.
}
\label{fig7_1}
\end{figure}

To comprehend the band structure shown in Fig.~\ref{fig7_1}, 
we also discuss the DOSs calculated from the TB model shown in Fig.~\ref{SOC_4}(a).
 We also find the same relationship of the eigenvalues as Fig.~\ref{fig1},
 $\epsilon^{\rm{SO}}_{\textsf{C}} < \epsilon^{\rm{SO}}_{\textsf{D}_v} < \epsilon^{\rm{SO}}_{\textsf{D}_c} < \epsilon^{\rm{SO}}_{\textsf{A}}$
with the insulating state 
 due to a finite gap around $\omega - \mu \simeq 0$. 
obtained using the first-principles DFT calculation.
 Both the behaviors are consistent with each other 
except for the region above $\omega = \mu$, 
where the width of the peak of the DFT calculation is larger. 

As described above, the inset of Fig.\ref{SOC_4}(a) denotes the DOSs calculated with the site-potential determined by this solution (2) in Sec.~\ref{sec:Opt_site}, where the reduction of DOS is enhanced for $\omega < \mu$. 
Here we mention the local density, which is estimated by Eq.~(\ref{eq:n}).
Using the site-potentials $\delta V_c$ = --0.043 eV ($\mu$ = 0.1637 eV ), calculated local charge density per spin is 
 $n_A = n_{A'}$= 1.46, $n_B$= 1.41, and $n_C$ = 1.67. 
With the compromise value of $\delta V_{C}$ = --0.03 eV, $n_A = n_{A'}$= 1.46, $n_B$= 1.42, and $n_C$ = 1.65 (Table~\ref{table2}). Note that 
both $n_A - n_B$ and $n_C -n_A$ increase 
 with decreasing  $\delta V_{C} ( <  0)$.   
In decreasing order by the local charge density, we find $n_C > n_A > n_B$, which agree with\ET.
Since these quantities are almost the same as those of Fig.~\ref{fig8},~the deviation of overall band structure due to the modification of site-potentials is small.  
However, we emphasize that electronic states close to the Dirac point ($<$ 0.015 eV) is modulated by the choice of the site-potential.
The present choice of the site-potentials for the SOC can be justified as a perturbation.  

\begin{figure}
  \centering
\includegraphics[width=4.5cm]{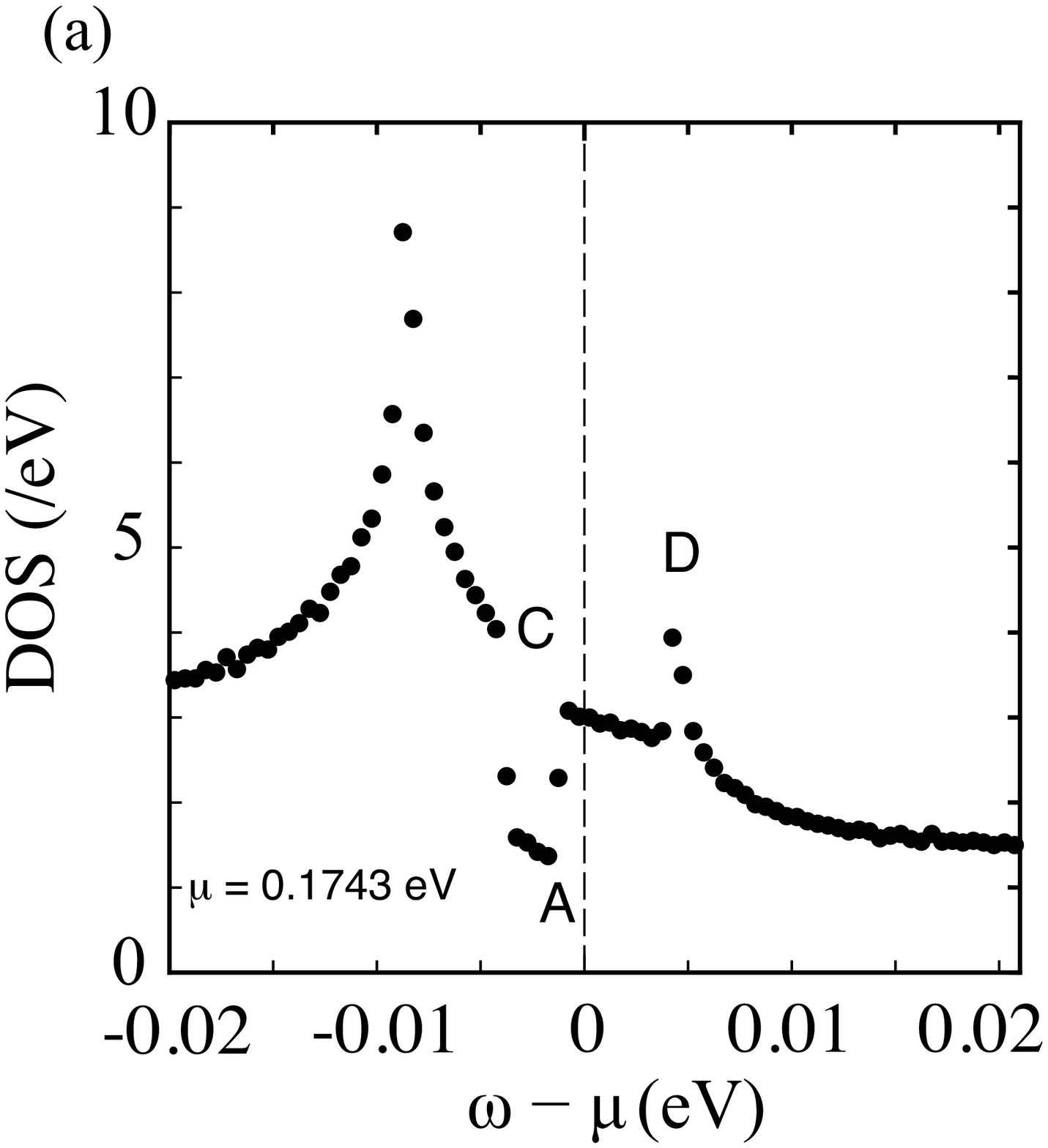} 
\includegraphics[width=4.2cm]{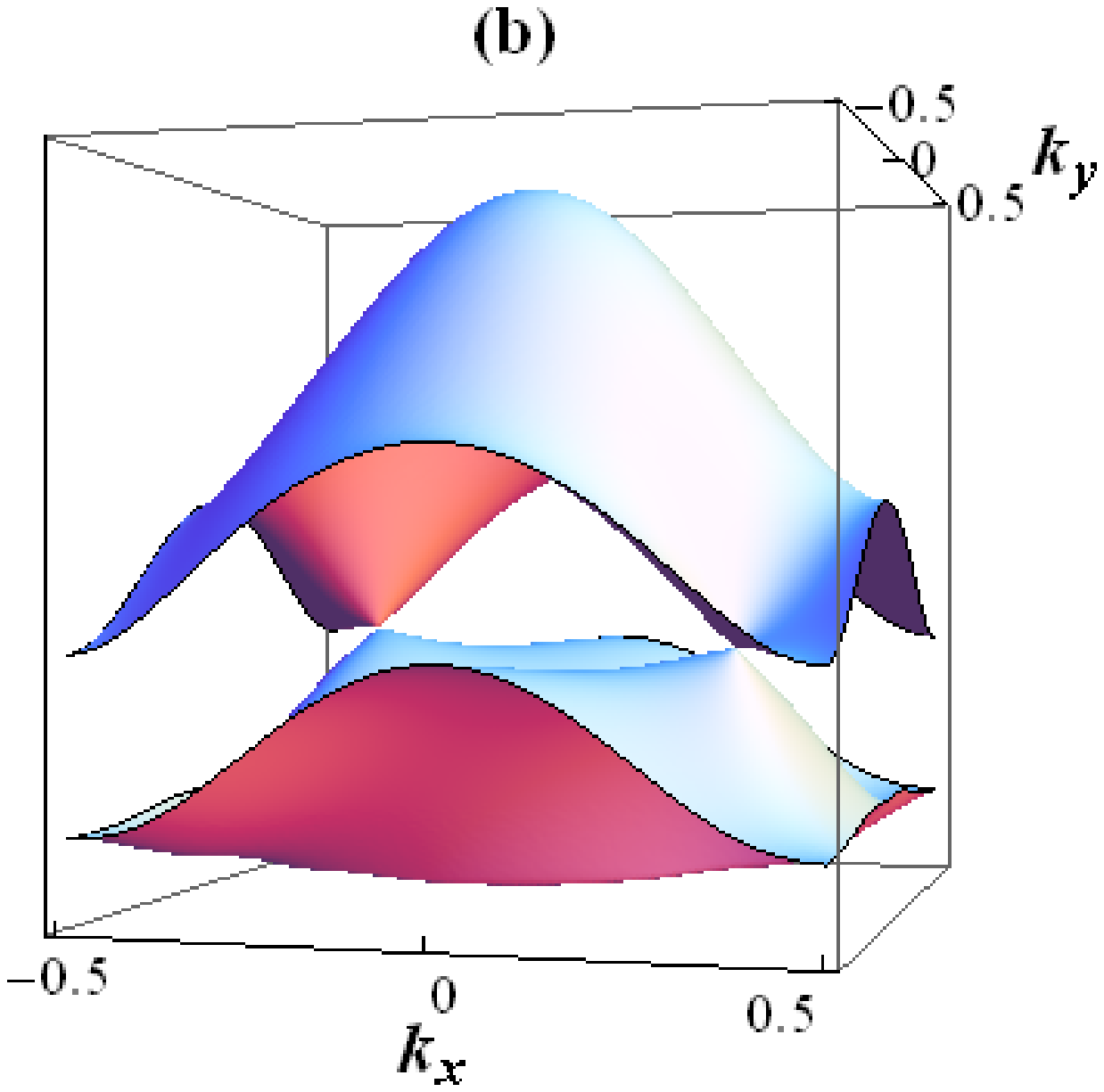} 
\caption{(Color online) 
(a) Density of states (DOS) without SOC obtained from transfer energies and site-potentials shown in Table~\ref{Transfer_alpha}. 
The chemical potential $\mu$ is located at 0.1743 eV. 
The relative energies of the eigenvalues of $\textsf{A}$, $\textsf{C}$, and $\textsf{D}$ with respect to $\mu$ are 
$\textsf{A}$ = --0.0013 eV, $\textsf{C}$ = --0.0039 eV, 
and $\textsf{D}$ = 0.0038 eV.
(b)~Two bands of $E_1(\bk)$ and $E_2(\bk)$ 
  for the TB model without SOC. 
  The Dirac point is given by 
 ${E_{1}(\bk)} = {E_{2}(\bk)}$ and is located 
 at $\bk$ = $\bkD$ = $\pm$(0.35, --0.30).
}
\label{fig7}
\end{figure}

\section{\label{sec:CompNoSOC} Non-SOC effective model and discussion}
~~~Next, we describe the results of an effective model derived from a non-SOC band structure, where the result with the DFT site-potentials 
  is shown in Fig.\ref{fig7}.    
The model parameters of the non-SOC case in Table~\ref{Transfer_alpha} provide
 the metallic state. 
 In fact, the Fermi surface exists on $E_2(\bk)$ suggesting a hole pocket around the Dirac point. 
 $E_1(\bk)$ reaches a local minimum at $\bkD$ and gives an electron pocket around the S (=M) point ($\textsf{C} < \textsf{A} < \textsf{D}$) as summarized in Table~\ref{table2}.
 We have shown that a precise effective Hamiltonian can be developed by using MLWFs generated from Bloch functions obtained in self-consistent full-relativistic DFT calculations, instead of those by scalar-relativistic calculations. 
 The magnitude of optimized potential for non-SOC, 
 which gives rise to the ZGS, 
   is too large and is beyond the present scheme.

Next, we discuss the origin of the difference in the relative position of eigenvalues (\textsf{A}, \textsf{C}, and \textsf{D}) between those without and with SOC (Fig.~\ref{fig7} and Fig.~\ref{fig8}, respectively).
As seen from Table~\ref{Transfer_alpha}, the SO transfer energies (i.e., the off-diagonal elements) are absent for the former band structure but present for the latter. 
However, interestingly, we found that the off-diagonal elements of SO transfer energies do not play a role in inverting the two energy positions ($\textsf{A}$ $<$ $\textsf{D}$ in Fig.~\ref{fig7} and $\textsf{D}$ $<$ $\textsf{A}$ in Fig.~\ref{fig8}).  
When we calculate a band structure including SOC by using transfer energies with only diagonal elements ($a_1$, $a_2$, $a_3$, $b_1$, $\cdots$ and $s_4$) in Table~\ref{Transfer_alpha}, the resultant two bands are almost the same as Fig.~\ref{fig8}(b) within the visible scale, where the obtained band structure maintains the relation of $\textsf{D}$ $<$ $\textsf{A}$. 
This implies that the relationship with $E_1$(S) in Fig.~\ref{fig8}(b) is determined by the diagonal elements rather than the off-diagonal elements. 
The diagonal elements contain a significant component coming from the mixing between different molecular orbitals which is originated from delocalized nature of Se $p$ orbitals,~\cite{TsumuP3HT, KuritaSOC} since such a SOC effect was not found in $\alpha$-(ET)$_2$I$_3$. 
Therefore, it turns out that the SOC  exhibits a remarkable role in 
 deriving  the effective 
  Hamiltonian of  $\alpha$-(BETS)$_2$I$_3$.

Last, we comment on the reliability of the present first-principles calculations. 
It has been reported that, when on-site Coulomb interaction $U$ are added to $p$ orbitals of S atom using GGA+$U$ method for quasi-one dimensional molecular conductors, the bandwidth tends to be narrower, indicating that the more localized behavior of the wave functions.\cite{AlemanyNMPTCNQ, KiyotaTTF2019}
Alternatively, a hybrid functional proposed by Heyd, Scuseria, and Ernzerhof (HSE06)\cite{HSE03,HSE06} also provides a proper description of insulating states in molecular solids.\cite{TTFCA_HSE, TsumuD3CatHSE}  
With this functional, the bandwidth and transfer energies generally increase, but the four bands near $E_F$ are farther apart from each other, which also indicates the localized nature. 
Moreover, distant transfer energies were also identified with HSE06 functional. 
We also note that in another molecular solid of the BETS molecule, distant transfer energies must be included to accurately reproduce the DFT band structure calculated with GGA-PBE. \cite{aizawaBETS}  
Therefore, we consider the enhancement of bandwidth and the existence of distant transfer energies to be intrinsic. 
A quantitative evaluation of the bandwidth and velocity with these methods will be performed and compared with experiments in the near future.

\section{\label{sec:level5}Conclusions}
~~~We proposed an effective model Hamiltonian for the Dirac electron in a quasi-2D molecular conductor of $ \alpha $-(BETS)$_2$I$_3$ at ambient pressure from first-principles calculations. 
In the presence of SOC, we found an insulating state with an indirect band gap of about 2~meV. 
The intermolecular transfer energies were obtained using MLWFs localized on BETS molecules. 
The model parameters for an exotic insulating state were derived from a self-consistent full relativistic DFT calculations. 
We have shown that SOC plays a remarkable role in deriving the effective Hamiltonian for $\alpha $-(BETS)$_2$I$_3$.
Compared with the electronic state of the sulfur analog of\ET, the bandwidth and transfer energies are generally large as is the number of relevant transfer integrals. 
However, the eigenvalues close to the Dirac points are in a quite-narrow energy window. 
Nonetheless, there are small but non-negligible energy differences between the DFT eigenvalues and those of a TB model. 
In order to reproduce the DFT bands with a moderate number of parameters, the inclusion of distant transfer integrals with small energies that we did not integrate into the TB model is essential. 
Therefore, we determined site-potentials that give the spectrum corresponding to DFT eigenvalues at several TRIMs by a reasonable fitting, and provide a reliable effective TB model for $\alpha$-(BETS)$_2$I$_3$.

\section{Acknowledgements}
~~~We thank H. Sawa, S. Kitou, A. Kobayashi, K. Yoshimi, D. Ohoki, K. Kishigi, F. Ishii, H. Sawahata, N. Tajima, S. Fujiyama, H. Maebashi, R.~Kato, M. Naka, and H.~Seo for fruitful discussions. 
This work was supported by a Grant-in-Aid for Scientific Research (Grants No. JP19K21860) and JST CREST Grant No. JPMJCR18I2. 
TT is partially supported by MEXT Japan, Leading Initiative for Excellent Young Researchers (LEADER).
Cooperative Research Program and the Supercomputing Consortium for the Center for Computational Materials Science at the Institute for Materials Research~(IMR), Tohoku University. 
The computations were mainly carried out using the computer facilities of ITO at the Research Institute for Information Technology, Kyushu University, and MASAMUNE at IMR, Tohoku University, Japan.
\section{Author contribution statement}
~~T.T. performed first-principles calculations, derived the
effective transfer energies, and wrote the manuscript. Y.S.
analyzed the effective tight-binding model.
Both T.T. and Y.S. agreed with all the contents of the present manuscript.
\appendix
\section{\label{sec:level_B}Site-energy potentials}
~~We define site-potentials acting on $B$ and $C$ sites, $\Delta V_{B}$ and $\Delta V_{C}$, which are measured from site-energy at $A$~($A^\prime$) site, $V_{A}$.\cite{Kondo2009}
\begin{subequations}
\begin{eqnarray}
\Delta V_{B} = V_{B} - V_{A},\nonumber \\
\label{siteVb}
\Delta V_{C} = V_{C} - V_{A},\nonumber
\label{siteVc}
\end{eqnarray}
\end{subequations}
where $V_{A}$, $V_{B}$, and $V_{C}$ are the site-energies at each molecule that were calculated using MLWFs $|\phi_{\alpha,0} \rangle$; 
\begin{eqnarray}
V_{\alpha} =\langle\phi_{\alpha, \sigma,0}| H |\phi_{\alpha, \sigma^{\prime},0} \rangle, \nonumber
\label{eq:eq3}
\end{eqnarray}
where ${\alpha}$ indicates $A$ (=~$A^{\prime}$), $B$, and $C$ molecules. 
These site-potentials are referred as to $\Delta V_{B}^{\rm{DFT}}$ and $\Delta V_{C}^{\rm{DFT}}$ in the present study, and listed in Table~\ref{Transfer_alpha}.

\section{Matrix elements} 
In terms of  Eq.~(\ref{equ1}) with    $X= \e^{i\kx}$, $\bar{X}=  \e^{-i\kx}$, 
    $Y= \e^{i\ky}$, and $\bar{Y}=  \e^{-i\ky}$, 
  matrix elements, $t_{ij} = (\hat{H})_{ij}$, are given by 
 \begin{eqnarray}
 t_{11}&=&  t_{22} = t_{55} =t_{66} =a_{1d}(Y+\bar{Y})
 +s1(X+\bar{X}) 
                          \; , \nonumber \\
 t_{33}&=& = t_{77}= a_{3d}(Y+\bar{Y})
     + s3(X + \bar{X})+ \Delta V_B
                           \; , \nonumber \\
 t_{44}&=&  t_{88} = a_{4d}(Y+\bar{Y})
  + s4(X + \bar{X})+ \Delta V_C 
                           \; , \nonumber \\
 t_{12}&=& a_3+a_2Y+d_0\bar{X}+ d_1XY 
                          \; , \nonumber \\
t_{13}&=&b_3+b_2\bar{X}+ c_2 \bar{X} Y + c_4\bar{X}\bar{Y}
                           \; , \nonumber \\
 t_{14}&=& b_4Y+b_1\bar{X}Y+c_1\bar{X} + c_3
                           \; , \nonumber \\
t_{23} &=& b_2+b_3\bar{X}+c_2 \bar{Y} + c_4 {Y}
                           \; , \nonumber \\
t_{24} &=& b_1+b_4\bar{X}+c_1Y + c_3\bar{X}Y
                           \; , \nonumber \\
t_{34}&=& a_1+a_1Y + d_2\bar{X}+d_3X 
                   + d_2 XY+d_3\bar{X}Y 
                                 \nonumber \\
t_{17} &=& b2_{so1} \bar{X} + c2_{so1}\bar{X} Y
                           + c4_{so1}\bar{X} \bar{Y}
                           \; , \nonumber \\
t_{18} &=&   b1_{so1}\bX Y  + b4_{so1} Y + c1_{so1} \bX 
                           \; , \nonumber \\
t_{27} &=& b2_{so1}  + c2_{so1} \bY + c4_{so1} Y 
                           \; , \nonumber \\
t_{28} &=& b1_{so1}  + c1_{so1}Y + c3_{so1}\bX Y 
                           \; , \nonumber \\
t_{35} &=& b2_{so2} X  + c2_{so2}X \bY + c4_{so2}X Y
                           \; , \nonumber \\
t_{36} &=& b2_{so2}  + c2_{so2} Y  + c4_{so2} \bY 
                           \; , \nonumber \\
t_{45} &=&  b1_{so2}X \bY  + b4_{so2}\bY  + c1_{so2}X
                           \; , \nonumber \\
t_{46} &=& b1_{so2}  + c1_{so2} \bY + c3_{so2}X  \bY 
                           \; ,  
\nonumber
\label{eq:tij}
\end{eqnarray}

 and
 $t_{15} = t_{16} = t_{25} = t_{26}
 =t_{37} = t_{38} = t_{47} = t_{48} = 0$, and 
 $t_{ji} = t_{ij}^*$.

-----------------------------------------------

\end{document}